\documentclass[amsmath,amssymb,mathfonts,superscriptaddress,prb,natbib,twocolumn]{revtex4-2}
\usepackage{graphicx}
\usepackage{color}
\usepackage{bm}
\usepackage[english]{babel}

\usepackage{amsfonts}
\usepackage{multirow}
\usepackage{makecell}
\usepackage{float}
\usepackage{comment}
\usepackage{enumitem}
\usepackage{verbatim}
\usepackage{mathtools}
\usepackage{esint}
\usepackage{tipa}
\usepackage{braket}
\usepackage{verbatim}
\usepackage{stackengine}

\usepackage[colorlinks]{hyperref}

\usepackage{bbm} 

\renewcommand{\textsubring}[1]{\ensurestackMath{\stackon[-12pt]{#1}{\mkern2mu\scriptstyle\circ}}}

\newcommand{\doublesubring}[1]{\ensurestackMath{\stackon[-12pt]{#1}{\mkern2mu\scriptstyle{\circ\circ}}}}
\newcommand{\subringleft}[1]{\ensurestackMath{\stackon[-12pt]{#1}{\mkern2mu\scriptstyle\circ{\blankchar}}}}
\newcommand{\subringright}[1]{\ensurestackMath{\stackon[-12pt]{#1}{\mkern2mu{\blankchar}\scriptstyle\circ}}}

\newcommand{\quadsubring}[1]{\doublesubring{\ensurestackMath{\stackon[-16pt]{#1}{\mkern2mu\scriptstyle{\circ\circ}}}}}
\newcommand{\doublesubringleft}[1]{\subringleft{\ensurestackMath{\stackon[-16pt]{#1}{\mkern2mu\scriptstyle\circ{\blankchar}}}}}

\newcommand{\doublesubx}[1]{\ensurestackMath{\stackon[-12pt]{#1}{\mkern2mu\scriptscriptstyle{\times\times}}}}
\newcommand{\subxleft}[1]{\ensurestackMath{\stackon[-12pt]{#1}{\mkern2mu\scriptscriptstyle\times\blankchar}}}
\newcommand{\subxright}[1]{\ensurestackMath{\stackon[-12pt]{#1}{\mkern2mu{\blankchar}\scriptscriptstyle\times}}}

\newcommand{\blankchar}{\color{white}{1}}

\begin{document}

\title{Long-range Entanglement and Role of Realistic Interaction in Braiding of Non-Abelian Quasiholes in Fractional Quantum Hall Phases}
\author{Ha Quang Trung$^{1}$, Qianhui Xu$^{1}$, and Bo Yang$^*$} 
\affiliation{Division of Physics and Applied Physics, Nanyang Technological University, Singapore 637371.}
\affiliation{Institute of High Performance Computing, A*STAR, Singapore, 138632.}
\email{yang.bo@ntu.edu.sg}

\date{\today}
\begin{abstract}
Parity conservation dictates that when fusing pairs of Moore-Read (MR) quasiholes, such that each pair of charge-$e/4$ anyons forms a charge-$e/2$ anyon, the parity of the numbers of $1$-anyon and $\psi$-anyon must be conserved within a given system. This idea is illustrated here using the Jack polynomial formalism, which also provides a basis to numerically study the dynamics of MR anyons. We find that parity conservation gives rise to a long-range ``entanglement" between statistics of one pair of anyons and the fusion channels of other anyons in the background. It is therefore important to account for all quasiholes in an experimental system in order to accurately predict the measurement outcome. We also examine the effect of two-body electron-electron interaction on the degeneracy of two anyon fusion channels. This understanding of the quasihole dynamics can help to fine-tune two-body interactions in order to stabilize the non-Abelian degeneracy, which is important for detecting non-Abelian braiding in experiments.
\end{abstract}

\maketitle

\section{Introduction}
The search for non-Abelian anyons is motivated by their promise of fault-tolerant quantum computation\cite{Freedman2000, Nayak2008,preskill1999lecture, sarma2015majorana, kitaev2003fault, Bigelow2019, kauffman2016braiding, kauffman2018majorana,alicea2011non, Kapit2012,feldman2021,yazdani2023hunting}. While anyons have been proposed in many theoretical models\cite{Leinaas1977,kitaev2003fault, alicea2011non,feldman2021,levin2005string, wilczek1990fractional}, experimental realizations are few. Only recently were the signatures of anyons (in particular, Majorana fermions) reported in novel platforms such as qubit systems\cite{xu2023experimenting,iqbal2024non} and topological superconductors\cite{nadj2014observation,Yasui2012, lv2017experimental, beenakker2020search}, although their true existence remains inconclusive\cite{frolov2021quantum}. Prior to these discoveries, realizations of anyons were studied extensively both theoretically and experimentally in fractional quantum Hall (FQH) systems - two-dimensional electron gas (2DEG) under an external magnetic field that exhibits topological order\cite{tsui1982two, wen1995topological,cage2012quantum}. Recent new experimental progress in quantum point contact experiments has enabled direct measurement of Abelian anyonic statistics\cite{bonderson2008interferometry, rosenow2016current, Nakamura2020,Bartolomei2020, ruelle2023comparing,lee2023partitioning,werkmeister2024anyon, samuelson2403anyonic} while constant efforts are being put in realizing their non-Abelian counterpart\cite{bonderson2006detecting, wu2014braiding,stern2006proposed,  Lee2022,willett2023interference}. In many systems currently accessible in the lab, braiding non-Abelian anyons and probing their statistics remain a challenge.

All different models of anyons share a similarity: one anyon is not a singular particle, but rather a low-energy collective excitation.  Ideally, they are realized by model Hamiltonians that are not readily accessible in real experiments. For example, Ising anyons in FQH phase are realized by the Moore-Read (MR) quasiholes at filling factor $\nu=5/2$\cite{moore1991nonabelions, nayak19962n,Nayak1996,Wan2008, Read1996,yazdani2023hunting}, which are exact zero-energy ground states of a three-body electron-electron interaction\cite{Wan2008,Prodan2009}. While it is believed that the $5/2$ plateau observed in the experiment is a non-Abelian phase, experiments probing the statistics of its quasiholes are complicated by the quasihole dynamics arising from Coulomb-type two-body interactions\cite{willett2023interference}. Previous studies\cite{Prodan2009,baraban2009numerical, Tserkovnyak2003,Bonderson2011, wu2014braiding, Macaluso2019} have highlighted several effects of Coulomb interaction on MR anyons: the splitting of the degeneracy required for non-Abelian statistics, the stability of the two fusion channels, etc. These studies, however, exhibit many caveats. For one, evaluating the Coulomb variational energy using the model wavefunction gives the physics in the lowest Landau level (LLL), whereas the MR state in experiments is realized on the second Landau level (1LL), whose properties have been shown to differ drastically from those of the LLL\cite{cage2012quantum, morf2002excitation,trung2021fractionalization}. Furthermore, many past studies, often constrained by numerical calculations, are limited in their scopes -- they study the energy spectra for only a handful of real-space configurations of anyons\cite{Prodan2009,Macaluso2019}, which may not be a reflection of the full anyon dynamics that vary as a non-monotonous function of their pair-wise separation\cite{trung2021fractionalization,xu2024}. An overall description of the microscopic mechanism for multiple fusion channels in non-Abelian anyons and the effects of realistic interaction is in general lacking.

Anyon fusion is the central rule in determining all physical properties of an anyon model, including mutual statistics\cite{bonderson2008interferometry, Macaluso2019}. In the context of FQH, each anyon is a quasihole excitation\footnote{A quasielectron in a FQH system is also anyonic, but our study does not concern quasielectrons because they require a finite energy even with respect to model Hamiltonian.}, parametrized by a location and possibly a quantum metric describing its shape\cite{arovas1984fractional,Umucalllar2018, Comparin2021, trung2023spin,nardin2023spin}. ``Fusing" two anyons simply means bringing their positions together so that they can be ``stacked" one on top of the other. In the context of Abelian braiding, this description gives rise to a microscopic derivation of the spin-statistics relation for anyons\cite{nardin2023spin,trung2023spin}. %In the MR state, fusing two $\sigma$ anyons (which are charge-$e/4$ quasiholes) results in two possible outcome, as signified by the fusion rule $\sigma\times\sigma=1+\psi$. Here $1$ and $\psi$ on the right-hand side denote two different species of quasiholes with charge $e/2$, $\times$ denotes the fusion process, i.e. bringing two charge-$e/4$ quasiholes together, and $+$ denotes linear combination of possible outcomes; the three symbols $\sigma$, $1$, and $\psi$ are called the topological charges of the MR anyons. 
In non-Abelian braiding, such as the MR phase, fusing two anyons may result in more than one outcome\cite{bonderson2006detecting}. The (approximate) energetic degeneracy of these two fusion channels under realistic electron-electron interaction is then necessary for non-Abelian braiding\cite{baraban2009numerical, Prodan2009,bonderson2009splitting, willett2023interference, Lee2022}. In the MR state the dynamics of two- and four-quasihole systems have been numerically studied in two limits: when quasiholes are very far from each other\cite{Prodan2009,storni2011localized}, and when two quasiholes completely fuse together\cite{baraban2009numerical}. 
However, at finite separation the energy gap may be finite as this degeneracy is not protected by any topology or symmetry. 
It is thus an important question as to how the electron-electron interaction can be tuned to close this gap and enable a platform for non-Abelian braiding, in systems where many quasiholes can be present and interacting with each other.

\begin{figure}
\begin{center}
\includegraphics[width=\linewidth]{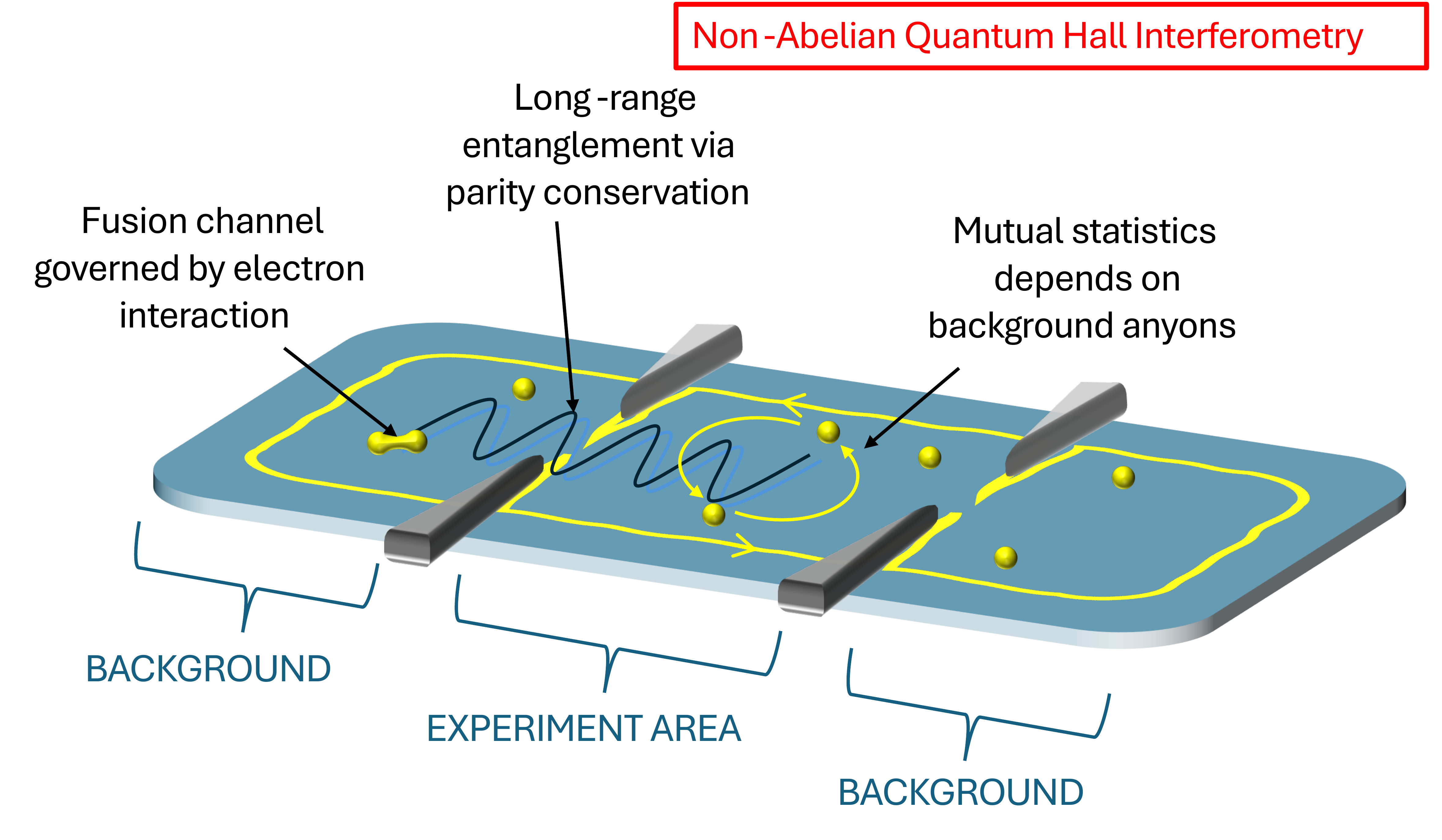}
\caption{If in the Moore-Read (MR) phase both charge-$e/4$ (yellow circles) and charge-$e/2$ (fused pair of circles) are present, the anyonic species of the charge-$e/2$ quasiholes will be determined by the underlying electronic interaction in the system. This in turn will affect the braiding measurement via a long-range entanglement originating from parity conservation.}
\label{fig:entanglement}
\end{center}
\end{figure}

Apart from energetics, the fusion channels of MR quasiholes are potentially dictated by another factor called parity conservation\cite{Macaluso2019, Baldelli2021}. %Theoretically speaking, parity conversation arises from the fusion rule $\psi\times\psi=1$. This fusion rule says that the number of $\psi$-anyon in a given system must remain either even or odd throughout all fusion processes. 
As will be explained in detail in the next section, two MR quasiholes can fuse in two different channels, labeled ``$1$'' and ``$\psi$''. The parity conservation places a constraint on the fusion outcome of the entire system, saying that the number of $\psi$ anyons in a given system always remains either even or odd. %This manifests as different statements of the ``even-odd effect"\cite{Macaluso2019,willett2023interference}.  In the case of the MR quasiholes, a
Aspects of this parity that have been studied include different edge mode countings in the ``odd-sector" and ``even-sector"\cite{ardonne2004filling}, as well as effects of fusion channels on Abelian braiding in the case of two quasiholes\cite{Macaluso2019,willett2023interference}. However, an important implication of the parity conservation that has not been pointed out and studied extensively is how the braiding of a set of anyons is influenced by not only their mutual statistics, but also the presence and fusion of any other anyons in the background \emph{even if the background anyons are very far away} (see Fig.\ref{fig:entanglement}).

In general, in braiding experiments on a FQH system, we are concerned with a general Hamiltonian of the following form:
\begin{equation}
\label{general Hamiltonian}
\hat H(t)=\lambda_0\hat H_{\text{model}} + \lambda_1\hat H_{2bdy} + v_0\hat H_{\text{pins}}(t)
\end{equation}
where $\hat H_{\text{model}}$ is the model Hamiltonian whose nullspace is the ground state and quasihole states of a certain FQH phase, $\hat H_{2bdy}$ is some two-body interactions (e.g. Coulomb interaction or Haldane pseudopotential), and $\hat H_{\text{pins}}$ is a one-body Hamiltonian consisting of some configuration of potential pins coming from e.g. the tip of a scanning tunneling microscope (STM) device\cite{papic2018imaging,jack2021detecting,hu2025high}. These potential pins trap the quasiholes in places, which is necessary for the manipulating them in real space to braid them. To braid two or more quasiholes one may vary their positions by tuning $\hat H_{\text{pins}}$ in time $t$, while $\hat H_{\text{model}}$ and $\hat H_{2bdy}$ typically remains constant in time. For non-Abelian phases, $\hat H_{\text{model}}$ typically involves three- or more-body interactions\cite{moore1991nonabelions}. The  $\lambda_1=0$ case presents an ideal scenario where each pin traps an ideal quasihole provided that $v_0\ll \lambda_0$ and the number of pins equals the number of quasiholes. The physics in this scenarios has been reported by several numerical studies in the past with regards to both the quasihole degeneracy in a finite system and braiding properties\cite{Prodan2009,baraban2009numerical, Tserkovnyak2003, wu2014braiding}. Adding the second term presents a potential challenge to numerical analysis using exact diagonalization (ED) as the Hamiltonian matrix becomes more dense\cite{Prodan2009}.

In this paper we introduce a general microscopic understanding of quasihole dynamics that can predict the effect of different terms in Eq.(\ref{general Hamiltonian}). Applying it to the context of the Moore-Read state, two-body interactions can be shown to mediate both the creation energy of a single anyon and interaction energy between them\cite{xu2024}. As a result, we show that different types of two-body interaction can strongly affect the entanglement between subsystems, and the effect is measurable in braiding experiments. We show how this understanding helps in tuning the two-body interaction to ensure ideal experimental conditions for detecting non-Abelian statistics.

This paper is organized as follows. In the next section we review the Ising anyon model and illustrate, using the Jack polynomial formalism\cite{bernevig2008model,bernevig2008generalized}, how the three types of Ising anyons manifest physically as MR quasiholes. We will also show that parity conservation can be seen from the root configurations of the Jack polynomials. Next, in Section III we discuss an important implication of parity conservation, which is a ``long-range entanglement'' that enables the fusion of one pair of anyons to directly affect the statistics of another pair elsewhere in the same system. In section IV, we numerically investigate the effects of two-body interactions on the energetics of MR quasiholes. This allows us to simulate the entanglement effect by demonstrating how simultaneous selection of fusion channel of two quasiholes in a four-quasihole MR system renders it Abelian with pre-determined statistics. We will also discuss methods to mitigate this effect to stabilize non-Abelian signatures in experiments.    
 
\section{Microscopic wavefunctions and parity conservation}
\subsection{Types of MR quasiholes}
Quasiholes of the MR states realize the Ising anyons\cite{bonderson2008interferometry}. In this model, there are three species of anyons, labelled by $\sigma$, $\psi$, and $1$ and they satisfy the fusion rules:
\begin{align}
\sigma\times\sigma&=1+\psi\label{fusion1}\\
\psi\times\psi&=1\label{fusion2}\\
\psi\times\sigma&=\label{fusion3}\sigma
\end{align}
In the MR state, $\sigma$-anyons are the charge-$e/4$ quasiholes, and $1$- and $\psi$-anyons are charge-$e/2$ quasiholes of different types. Here we explicitly construct the microscopic wavefunctions for these three types of anyons using the Jack polynomial formalism. \cite{bernevig2008model, bernevig2008generalized}. This approach exhibits two major advantages, one conceptual and one practical. Conceptually, it provides a simple understanding of the parity conservation in terms of ``root configurations". Practically, the Jack polynomials provide a basis for obtaining via ED the low-lying states of Hamiltonian of the form given by Eq. (\ref{general Hamiltonian}), in the case where $\hat H_{\text{model}}=\hat V_3^{3bdy}$ is the three-body interaction and $\lambda_0\to\infty$. This enables us to greatly reduce the dimension of the Hilbert space involved, but with a small price that the Hamiltonian matrix is no longer sparse\cite{Prodan2009}. Throughout this paper, we work on the spherical geometry\cite{haldane1983fractional}.

\begin{figure}
    \centering
    \includegraphics[width=\linewidth]{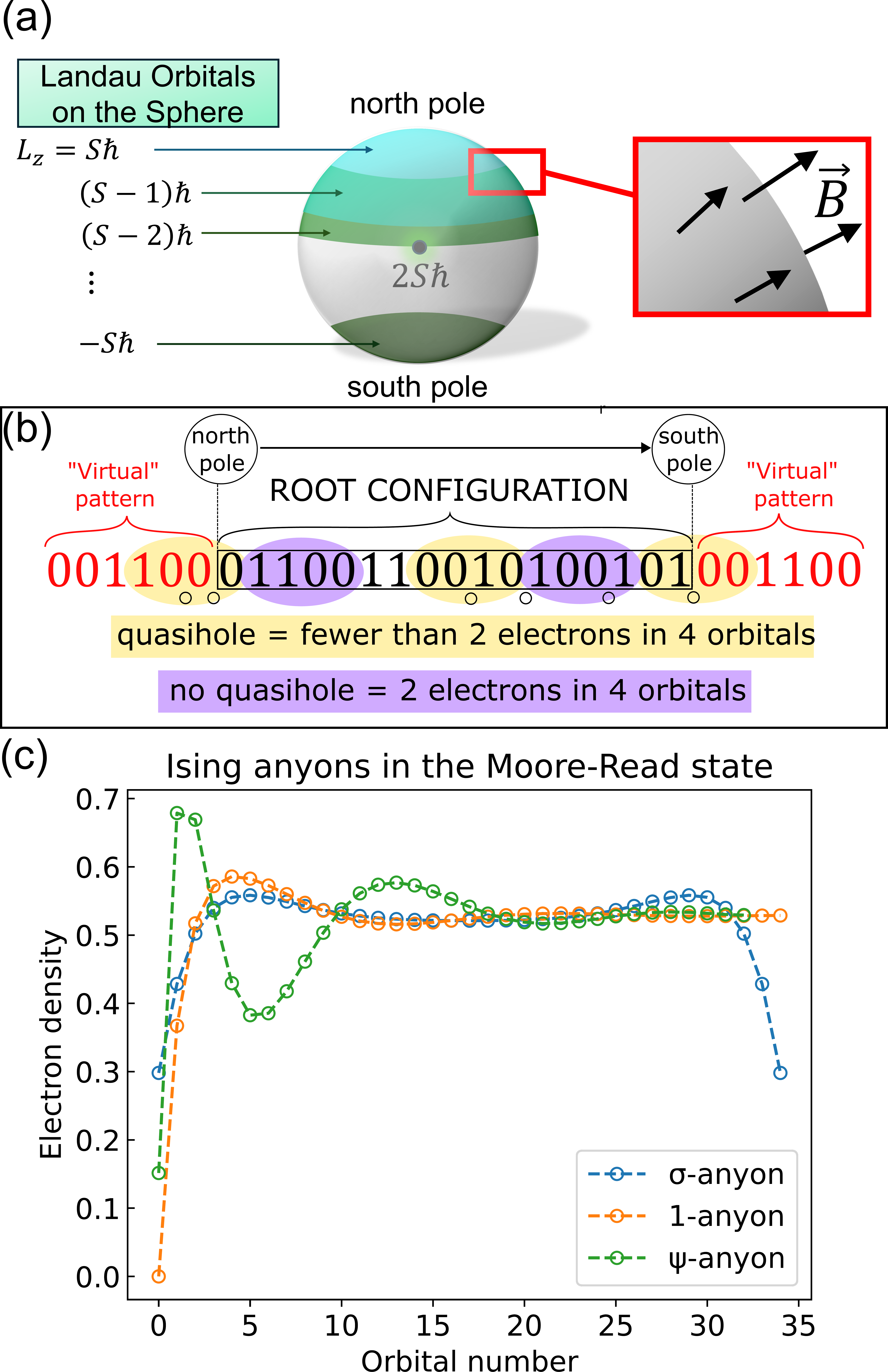}
    \caption{(a) The spherical geometry used in this paper\cite{haldane1983fractional}: a magnetic monopole of strength $2S\hbar$ (where $S$ is either integer or half-integer) is placed at the center of a sphere, creating a uniform magnetic field everywhere perpendicular to the surface of the sphere. The Landau orbitals are concentric rings parallel to the equator\cite{greiter2011landau}(b) A root configuration is a binary string with each digit representing an orbital from the north pole to the south pole going right to left. To read off the position of MR quasiholes, every group of four consecutive digits is checked (examplified by yellow and purple ellipses) and a quasihole is marked whenever there are fewer than two electrons in it (examplified by yellow ellipses). The root is extended on the two sides by ``virtual'' repetitions of ``1100'' in order to read off the quasiholes residing at the two poles. (c) The density per orbital on the sphere showing the three types of anyons in the Moore-Read state: two $\sigma$-anyons at each pole (blue); one $1$-type at the north pole (orange); and one $\psi$-type at the north pole (green).}
    \label{fig:MR quasiholes}
\end{figure}

Following the Jack polynomial formalism, the ground state and quasihole wavefunctions of the Moore-Read state are the Jack polynomials with root configurations that satisfies the ``$(2,4)$-admissibility rule". A root configuration can be represented by a string of digits 0's and 1's, and the $(2,4)$-admissible rule says that any group of four consecutive digits contains no more than two digit 1's. Physically, each digit is interpreted as a Landau orbital, where ``1" means an electron is occupying the corresponding orbital and ``0" means the orbital is empty. On the spherical geometry, the left end of the root configuration also corresponds to the orbital at the north pole while the right end corresponds to the south pole (see Fig. \ref{fig:MR quasiholes}). As a result of the $(2,4)$-admissible rule, a root with $N_e$ electrons must have at least $2N_e-2$ orbitals in total. There exists a unique root with $N_e$ electrons and $N_o=2N_e-2$ orbitals:
\begin{equation}
\label{MR ground state root}
1100110011...110011
\end{equation}
where ``..." denotes a pattern of repeating ``$1100$". This root correspond to the MR ground state, whose first-quantized wavefunction is the well-known Pfaffian:
\begin{equation}
\label{MR ground state wavefunction}
\psi_{MR}(z)\propto \text{Pf}\left(\frac{1}{z_i-z_j}\right)\prod_{i<j}(z_i-z_j)^2
\end{equation}

Quasiholes can be added to the ground state by insertion of additional orbitals. Symbolically this is done by adding 0's to the root configuration. One example is:
\begin{equation}
    \label{eq:MR 1qh}
    \doublesubring{\blankchar}01100110011...110011
\end{equation}
where we use the empty circles below the root configuration to denote the positions of the quasiholes. Quasiholes are located in the root configuration where four consecutive orbitals contain fewer than two electrons (see Fig. \ref{fig:MR quasiholes}b)\cite{seesup}.  Note that while this is seemingly a tongue-in-cheek method to mark down quasihole positions, there exists an empirical correspondence between the orbital positions marked in the root configuration and the deficiency in electron density in the actual state obtained from the Jack polynomial parametrized by that root\cite{yang2013analytic,yang2014nature,trung2021fractionalization}. This is a special aspect that makes the Jack formalism especially useful in studying the physics of quasiholes\cite{trung2021fractionalization,xu2024}.

The 0's and 1's in the root configuration shown in Eq. \ref{eq:MR 1qh} can be shuffled such that the two quasiholes are separated:
\begin{equation}
    \label{eq:MR 2qh}
    \subringleft{1}010101...1010\subringright{1}
\end{equation}
where the ellipsis now denotes repeating patterns of ``10". There exists another configuration where two quasiholes are closed together, such that they form a bound pair:
\begin{equation}
\label{eq:MR 1qh psi highest weight}
\doublesubring{1}00110011...110011
\end{equation}
Eq.(\ref{eq:MR 1qh})-(\ref{eq:MR 1qh psi highest weight}) represent the three types of Ising anyons present in the MR state: the $\sigma$-anyon is one empty circle in Eq.(\ref{eq:MR 2qh}), while the $1$-anyon and $\psi$-anyon refer to the bound pairs in Eq.(\ref{eq:MR 1qh}) and (\ref{eq:MR 1qh psi highest weight}) respectively. An empirical observation is that between quasiholes that makes up an $1$-anyon there are an even number of electrons in the root configuration (0 in Eq.(\ref{eq:MR 1qh}), where as that number is odd for the $\psi$-anyon (1 in Eq.(\ref{eq:MR 1qh psi highest weight})). This is a handy way to read off the fusion outcome of two $\sigma$-anyon in any given root configuration. 

We would also like to emphasize an observation about the root configuration for the $\psi$-anyon in the MR state, first reported in Ref. \cite{trung2023erratum}. Although Eq.(\ref{eq:MR 1qh psi highest weight}) gives a highest-weight state, and in our symbolic representations the two circles are side-by-side, the electron density at the center of the quasihole is actually quite high\cite{trung2023erratum}. There exists another root configuration which gives a lower electron density at the north pole:
\begin{equation}
    \label{eq:MR 1qh psi}
    \subringleft{1}010\subringright{1}00110011...110011
\end{equation}
If one applies a single one-body potential pin at the north pole on a MR system with an odd number of electrons (denoted as $N_e$) and $2N_e-1$ orbitals, then the ground state will be exactly the Jack given by root configuration in Eq.(\ref{eq:MR 1qh psi}) and not Eq.(\ref{eq:MR 1qh psi highest weight}). Since our analysis in this paper relies on the interplay between one-body and two-body interaction, we will use Eq.(\ref{eq:MR 1qh psi}) instead of Eq.(\ref{eq:MR 1qh psi highest weight}) as the model state for the $\psi$-anyon. The electron densities of the wavefunctions on the sphere corresponding to the three types of anyons are shown in Fig.\ref{fig:MR quasiholes}.

Note that the total number of electrons in Eq.(\ref{eq:MR 1qh psi}) is odd whereas it is even in Eq.(\ref{eq:MR 1qh}) and (\ref{eq:MR 2qh}). This reflects the fact that with only one added orbital to the ground state, the $1$-anyon and $\psi$-anyon do not mix and the fusion outcome of any two $\sigma$-anyons is unique. Thus a system with only two MR quasiholes can only exhibit Abelian statistics, while non-Abelian statistics only manifest when there are four or more quasiholes.

\begin{figure}
    \centering
    \includegraphics[width=\linewidth]{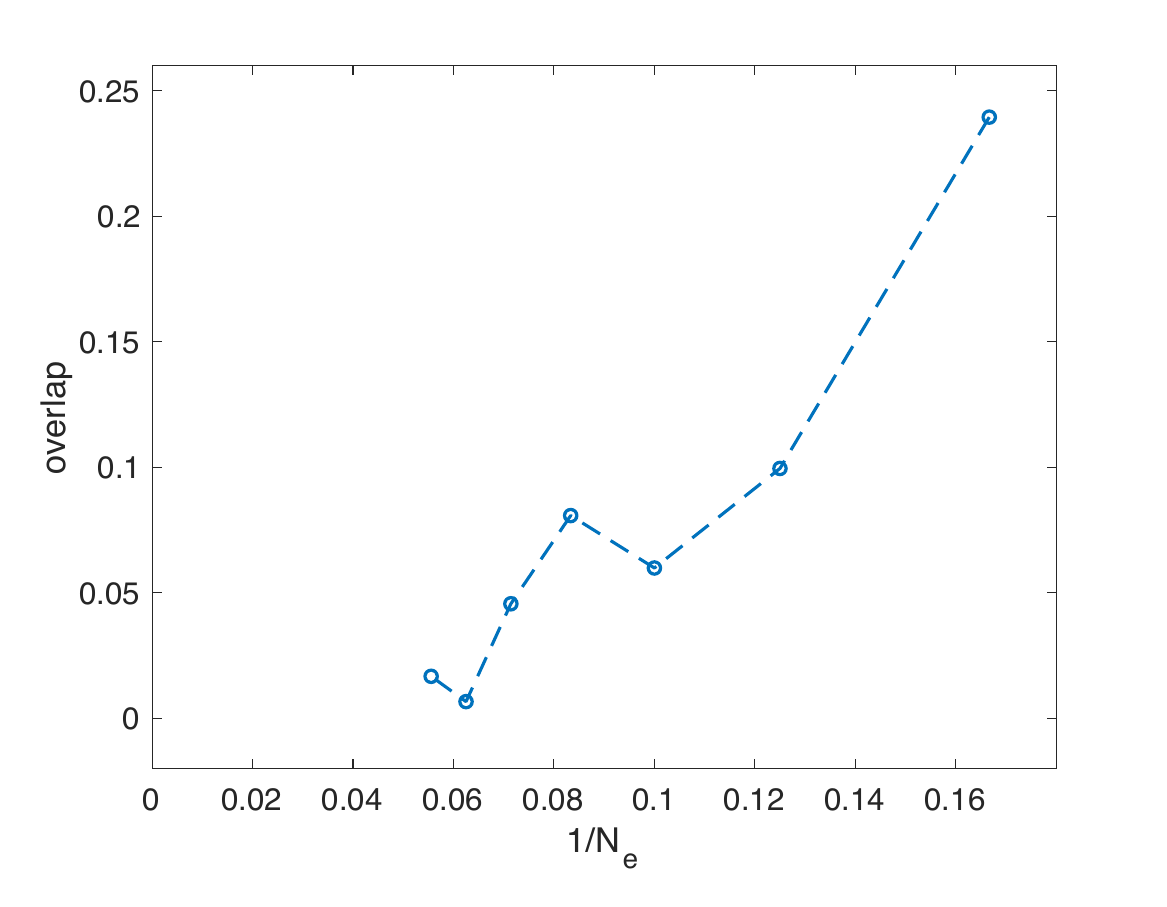}
    \caption{Finite-size scaling of the overlap of the two Jacks describing different types of MR 2-stack quasiholes. Their roots are given by Eq.(\ref{two 1-types}) and Eq.(\ref{two psi-types}). Maximum system size used in calculation has $N_e=18$ electrons.}
    \label{fig:overlap}
\end{figure}

\subsection{Parity conservation in systems with four or more quasiholes}
Given a fixed number of electron $N_e$, one may study different possible root configurations when two additional orbitals are added. If $N_e$ is even, a few such configurations are listed below:
\begin{align}
    \textsubring{\blankchar}\textsubring{\blankchar}\doublesubring{0}011001100110011...110011\label{eq:MR root1}\\
    %\textsubring{\blankchar}\doublesubring{0}10\subringright{1}00110011...110011\label{eq:MR root2}\\
    \doublesubring{\blankchar}0110\doublesubring{0}01100110011...110011\label{eq:MR root3}\\
    \subxleft{1}010\subxright{1}00\doublesubx{1}00110011...110011\label{eq:MR root4}\\
    \subxleft{1}010\subxright{1}00\subxleft{1}010\subxright{1}0011...110011\label{eq:MR root5}\\
    \vdots\hspace{2cm}\nonumber
\end{align}
where in Eq.(\ref{eq:MR root4}), we changed the empty circles into crosses to denotes the pair of quasiholes that fuses into a $\psi$-type anyon\footnote{Two MR quasiholes are denoted as empty circles if there are an even number of digit 1's between them, and as crosses if there are an odd number of 1's between them.}. This change of notation serves to distinguish different anyon species within the same system and makes it easier to observe the parity conservation. In this system with two pair of quasiholes, one can read off the fusion result of each pair of quasiholes (two consecutive empty circles) based on the number of electrons (digit 1's) between them. Two quasiholes fuse into a $1$-anyon if there are an even number of electrons between them, and $\psi$-anyon if there are an odd number of electrons between them -- this can also be observed in Eqs. (\ref{eq:MR 1qh}) and (\ref{eq:MR 1qh psi}). In Eqs. (\ref{eq:MR root1}) and (\ref{eq:MR root3}) there are two $1$-anyons and in Eqs. (\ref{eq:MR root4}) and (\ref{eq:MR root5}) there are two $\psi$-anyons. One can now observe that in a given root configuration a fixed number of orbitals and electrons, two $1$-anyons may be transmuted into two $\psi$-anyons and vice versa by shifting the electrons around, but the \emph{parity} of the number of $\psi$-anyons (its oddness or evenness) is a conserved quantity. 

Following the patterns of Eq.(\ref{eq:MR root1})-(\ref{eq:MR root5}), one can try to completely separate the two charge-$e/2$ quasiholes so that one is isolated at the north pole while the other at the south pole. There are two ways to do so, which corresponding to either a pair of $1$-types:
\begin{equation}
    \label{two 1-types}
    \doublesubring{0}1100110011...1100110011\doublesubring{0}
\end{equation}
 or a pair of $\psi$-type:
\begin{equation}
    \label{two psi-types}
    \subxleft{1}010\subxright{1}001100...001100\subxleft{1}010\subxright{1}
\end{equation}
Note that although for a finite system, the wavefunctions corresponding to Eq.(\ref{two 1-types}) and Eq.(\ref{two psi-types}) are \emph{not} orthogonal, their overlap decreases rapidly with increasing system sizes (see Fig.\ref{fig:overlap}). This indicates that these are indeed two different types of anyons. The finite overlap at finite system size can be attributed to the overlapping tails of the anyons due to finite separation.

In the above example we started with an even number of electrons and observe that we can have either none or two $1$-anyons. If we start with an odd number of electrons instead, then the system always have one $1$-anyon and one $\psi$-anyon, as illustrated by the following root configuration:
\begin{equation}
    \label{eq: MR root5}
    \doublesubring{\blankchar}0\doublesubx{1}00110011...110011
\end{equation}
More generally, in a system containing $2n$ MR quasiholes with $N_e$ electrons, then the even-$N_e$ and odd-$N_e$ cases are topologically distinct sectors, distinguished by different parity of the number $\psi$-anyons. Using our circles and crosses notation as above, one can check by inspection that the number of crosses is always $2k$ ($0\leq k\leq n$) where $k$ is an even integer if $N_e$ is even and odd integer if $N_e$ is odd. This means that a system with an even number of electrons can only have 0,2,4,... pairs of quasiholes that fuse into $\psi$-type anyon (any root configuration must have 0,4,8,... crosses) while a system with an odd number of electrons must have 1,3,5,... $\psi$-type pairs (corresponding to 2,6,10,... crosses in the root configurations). %The Hilbert space structure in each of these sectors is characterized by the even- and odd-sector countings of the partition functions in CFT{\color{red} you never mentioned what CFT is anywhere else}. 

%Finally, the root configurations described above can be used to construct real microscopic states. Working on the spherical geometry, one may view the left-most orbital of each root as the north pole and the right-most orbital as the south pole. Then, the real-space density of each state in Eq.(\ref{eq:MR 1qh})-(\ref{eq:MR 1qh psi}) can be calculated, which shows the qualitative difference in local electron density of each type of anyon (see Fig. \ref{fig:MR quasiholes}).

\section{Simultaneous selection of fusion channels}

%{\color{red} this is the most interesting section. it is probably better to put it as section 3, proposing all the physical ideas, and remove any numerics. the numerics itself should be an entire section. the energetics of MR quasiholes can be its subsection} 
%Apart from determining whether the non-Abelian degeneracy of the system is broken, the energetics of quasiholes also play an important role in dictating the braiding statistics of $\sigma$-quasihole. At first glance, this seems rather surprising. One may naively expect interaction should have little effect on the behaviors of $\sigma$-quasiholes we are trying to braid, as long as they are sufficiently far apart from each other. Indeed, if the system consists only of $2n$ well-separated $\sigma$-quasiholes (for some integer $n$), then every state in the $2^{n-1}$-fold degenerate ground state has exactly the same energy regardless of the electron interaction\cite{xu2024}. However, if some pairs of the quasiholes are fused, i.e. the MR systems contains both charge-$e/4$ and charge-$e/2$ quasiholes, then the types of the charge-$e/2$ quasiholes, which is determined by the two-body interactions, will influence the statistics of the charge-$e/4$ quasiholes. 

Parity conservation places a constrain on the possible outcome of a fusion of two MR quasiholes. To take a simplest example, let us consider the four-quasihole MR state in the even sector. If one pair of quasiholes is fused with a pre-determined outcome, let's say, into a $1$-anyon quasihole, then the fusion out come of the other pair is also known. Now, interesting physics can happen due to the following observation: \emph{the fusion outcome of two quasiholes directly affects their mutual statistics}\cite{Macaluso2019,nardin2023spin,trung2023spin}. This means that by pre-determining the fusion outcome of one pair of quasiholes, one can directly influence the statistics of other pairs of quasiholes in the system \emph{regardless of the distance between them}. Thus, parity conservation acts like a long-range entanglement between pairs of MR quasiholes (see Fig. \ref{fig:entanglement}).

In this section, we explore the physics resulting from this ``parity-mediated entanglement''. We start this section by a simple analogy between parity conservation in the MR state and typical picture of entanglement in quantum mechanics (i.e. the Bell state). This will motivate discussions on the effect of spontaneous fusion channel selection on braiding measurement, which we discuss subsequently. Numerical evidences for this effect will also be presented in the next section.

\subsection{Parity conservation as long-range entanglement}
We first provide here a short discussion on how parity can be seen as an entanglement. We draw analogy to the famous example of entanglement: the Bell state consisting of two electrons:
\begin{equation}
\label{Bell state}
|\psi\rangle = \frac{1}{\sqrt{2}}\left(|\uparrow\downarrow\rangle+|\downarrow\uparrow\rangle\right)
\end{equation}
This state has the remarkable property that measuring the spin of one electron (which consequently collapse the wavefunction according to the outcome) immediately determines the spin of the other. A simple way to view this entanglement is by looking at the unconditional and conditional probability of measuring up spin in the first electron:
\begin{align}
P(S_1\uparrow) &=1/2\\
P(S_1\uparrow | S_2\uparrow) &=0
\end{align}
where $S_i\uparrow$ denotes the outcome of measuring spin up in the $i$-th electron. Entanglement in this case is evidence in the fact that ``$S_1\uparrow$'' and ``$S_2\uparrow$'' are not independent events.

Coming back to the MR system, suppose we have four MR quasiholes that are separated into two pairs. Each pair, on its own, can fuse into either $1$ or $\psi$. Due to parity conservation, however, the two pairs must both fuse into $1$ or into $\psi$ (here we first take the even sector as an example), so the two possible outcomes can be written as $|\psi,\psi\rangle$ or $|1,1\rangle$. One can then construct a state similar to Eq.(\ref{Bell state}) as:
\begin{equation}
\label{Bell MR state}
\frac{1}{\sqrt2}\left(|\psi,\psi\rangle+|1,1\rangle\right)
\end{equation}
Similar to what happens in the Bell state, here the fusion outcome of one pair of anyons, if known, uniquely determines the fusion outcome of the other regardless of their spatial separation. Note that one can similarly construct an entangled state in the odd sector, where there is exactly one $\psi$-anyon and one $1$-anyon:
\begin{equation}
\label{Bell MR state odd}
\frac{1}{\sqrt2}\left(|1,\psi\rangle+|\psi,1\rangle\right)
\end{equation} 

Constructing model ``Bell-type'' states in this manner unveils the mechanism with which parity conservation can affect the physics of a subsystem. Let us next consider a MR system with six quasiholes. In this case, there can be up to three $\psi$-anyons. In the even sector, there can either be zero or two $\psi$-anyons, and in the odd sector there can be either one or three. Without loss of generality, let us take the even sector and consider the following state:
\begin{equation}
\label{Bell MR state 6 qh}
\frac{1}{2}\left(|\psi,\psi,1\rangle+|\psi,1,\psi\rangle+|1,\psi,\psi\rangle+|1,1,1\rangle\right)
\end{equation}
which is a possible state with six quasiholes forming three pair, each of which may fuse into either $1$ or $\psi$, but the total number of $\psi$ must be even. We partition the quasiholes into two parts: a ``background'' part consisting of the first pair (whose fusion outcome is denoted by the first symbol in each ket in Eq.(\ref{Bell MR state 6 qh}), and a ``subsystem'' part consisting of the second and third pairs (whose fusion outcomes are denoted by the second and third symbols, respectively). By using the simple calculation of conditional probability as illustrated above, one can verify that the background and subsystem are indeed entangled. For example, suppose we compute the probability of both the second and third pairs fusing into 1:
\begin{align}
P(1_{(2)},1_{(3)})&=1/4\\
P(1_{(2)},1_{(3)}|\psi_{(1)})&=0
\end{align}
(Here we use the notation $\alpha_{(i)}$ to mean the outcome ``the $i$-th quasihole pair fuses into one $\alpha$-type quasihole''.)

Now, suppose we make a ``measurement'' of the fusion outcome of the first pair of quasiholes. Realistically, this is done by bringing them together and tuning the interaction such that one fusion channel is energetically favoured over the other (the exact details are presented in the next section). In that case, we see that Eq.(\ref{Bell MR state 6 qh}) collapses into either of the following:
\begin{align}
\frac{1}{\sqrt2}|1\rangle\otimes\left(|\psi,\psi\rangle+|1,1\rangle\right)&&\text{if outcome is $1$}\label{collapse1}\\
\frac{1}{\sqrt2}|\psi\rangle\otimes\left(|1,\psi\rangle+|\psi,1\rangle\right)&&\text{if outcome is $\psi$}\label{collapse2}
\end{align}
Comparing these two with Eqs. (\ref{Bell MR state}) and (\ref{Bell MR state odd}), we see that the state describing the ``subsytem'' reduces to one of the two states describing four quasiholes. Interestingly, where as previously Eqs. (\ref{Bell MR state}) and (\ref{Bell MR state odd}) belong to topologically distinct Hilbert spaces and never mixes, here both Eqs. (\ref{collapse1}) and (\ref{collapse2}) reside within the same Hilbert space. More strikingly is the fact that by fusing a pair of anyons in background and tuning the interaction such that the outcome is $\psi$, we now have a subsystem in Eq. (\ref{collapse2}), which resembles the odd-sector four-quasihole state, within a system with even parity. We will argue that this ``parity flip'' of a subsystem has a measurable consequence in the anyonic statistics.

We note that all entangled states can only exist within the degenerate nullspace of the parent model Hamiltonian of the MR state, i.e. the three-body interaction $\hat V_{3}^{3bdy}$. Under some two-body interaction, the groundstate degeneracy is split, and Eq.(\ref{Bell MR state}) collapses into either of each component depending on whether $1$-anyon or $\psi$-anyon are energetically favorable. The exact details on the energetics of different anyon species will be discussed in the following section. %In this section, we focus on the effect of fusion channel selection of one pair of quasihole on the braiding statistics of the other, and argue that between the two pairs there exists a long-range entanglement mediated by parity conservation.
%The above discussion on entanglement missed out the fact that in the MR system, there are two qualitatively different ways in which the fusion channel of a pair of quasiholes may be determined experimentally: by bringing them together and obseriving the outcome, and by measuring their mutual statistics. Assuming the every pair of quasiholes are infinitely far away from each other, the second method does not intrinsically favor one outcome over the other even in the presence of two-body interactions. If we start with an initial state as Eq.(\ref{Bell MR state}), then the first method results in the wavefunction ``collapsing" into either of the two terms, while the second method resulting in different linear combination of the two terms given by the braiding matrix associated with that specific process.

\subsection{Effects on braiding statistics}
%The degeneracy of two fusion channels for MR quasiholes plays a decisive role in determine the outcome of the braiding processes. This degeneracy has been a subject of intensive studies in the past\cite{baraban2009numerical,Prodan2009,storni2011localized}. Ref. \cite{baraban2009numerical} showed that under Coulomb interaction, the $\psi$-anyon channel is slightly energetically favored. However, comparing the local density profile of the $1$-anyon and $\psi$-anyon, one may expect that a very sharp potential pin will slightly favor the $1$-anyon. Thus, an exact degeneracy may be achieved by a very delicate balance between the strengths of Coulomb interaction and potential pins. %However, real experimental conditions are more complicated and a more systematic study is required.

\begin{figure}
    \centering
    \includegraphics[width=\linewidth]{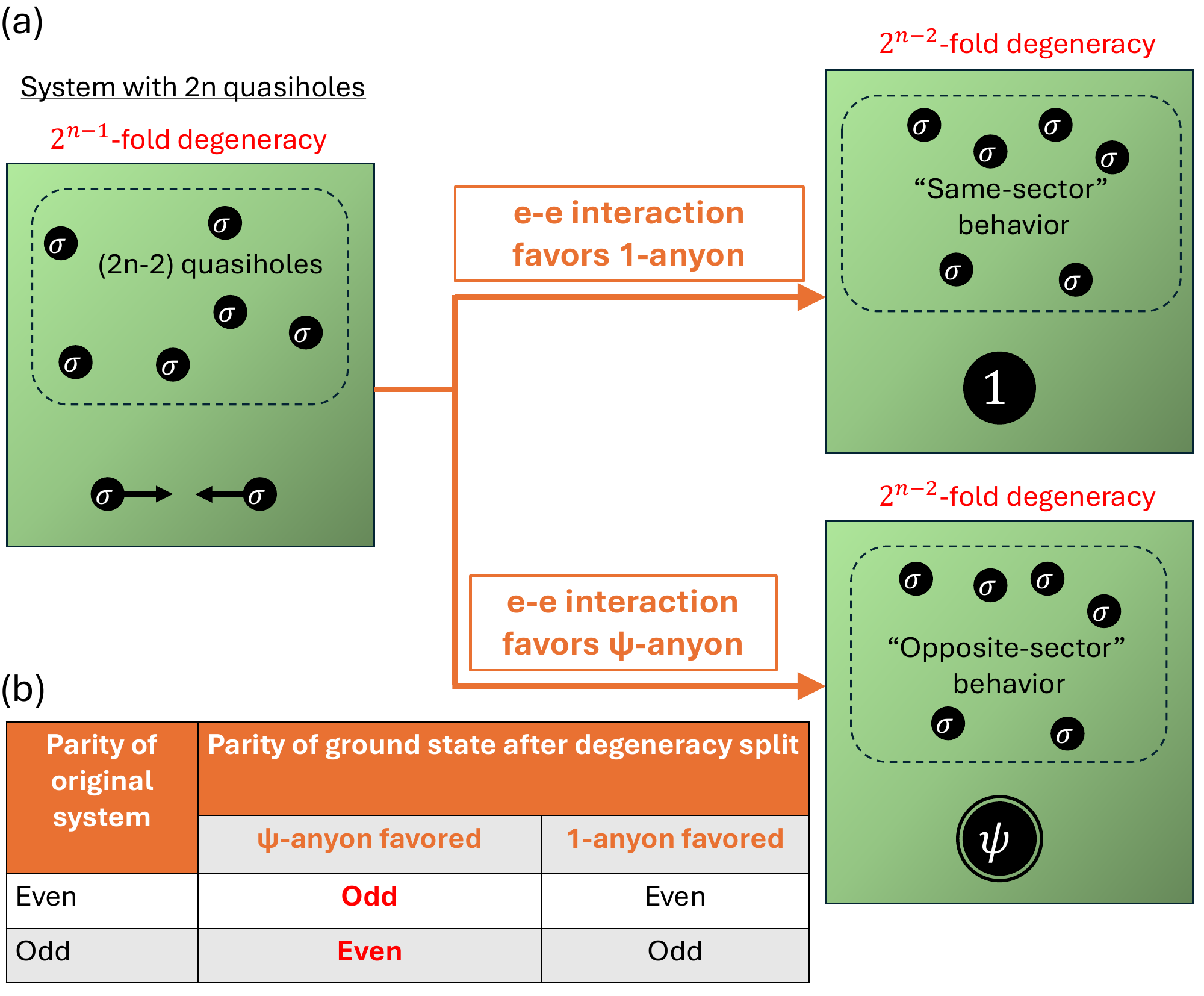}
    \caption{If the local measurements (dashed box) do not account for all quasiholes in the system, the behavior of the local system may differ from the global one depending on the effective electron-electron interaction (a) Illustration of the degeneracy-splitting mechanism (b) Relation of the parity of the low-lying states after the degeneracy splitting with that of the original system. Highlighted in red are the cases where the parity of the low-lying manifold is flipped after degeneracy splitting.}
    \label{fig:MR degeneracy splitting}
\end{figure}

In a FQH system there are two separate ways the fusion outcome of two anyons can be realized and measured: by placing two anyons at the same spatial co-ordinate, effectively ``fusing'' them in real space, or by measuring their mutual statistics via braiding. Assuming that during the braiding process the separation between every pair of $\sigma$-anyons remains very large (compared to their sizes), there exists a qualitative difference between the two methods. Namely, the presence of two-body interaction does not affect the degeneracy of the system during the braiding process, %(again, assumming only $\sigma$-anyon are present, and they are all well-separated from each other), 
but two-body interaction \emph{does} split the degeneracy when two $\sigma$-anyons are fused in real space. In the latter case, the fusion channel is chosen energetically depending on the details of two-body interaction (see Fig.\ref{fig:MR degeneracy splitting}). %{\color{red} again the transition from this to next section is not very good}

Interesting observations can then be made if we perform different ways of measuring the fusion outcome on different pairs of quasiholes within the same system. Let us consider a Hamiltonian of the following form:
\begin{equation}
    \label{eq:Hamiltonian}
    \hat H = \lambda_0\hat V_{3}^{3bdy}+\lambda_1\hat V_1^{\text{model}}+\lambda_2\hat V_\psi^{\text{model}} + \hat V_{\text{pins}}
\end{equation}
which resembles Eq. (\ref{general Hamiltonian}) where $\hat V_{3}^{3bdy}$ is the three-body interaction the nullspace of which is the Moore-Read quasihole manifold, $\hat V_\psi^{\text{model}}$ is some ``model'' two-body interaction that energetically favors $\psi$-anyon over $1$-anyon, and $\hat V_1^{\text{model}}$ is some other ``model'' two-body interaction that favors $1$-anyon over $\psi$-anyon, and $\hat V_{\text{pins}}$ is a collection of some local one-body pinning potentials. In the next section we shall see that indeed $\hat V_1^{\text{model}}$ and $\hat V_\psi^{\text{model}}$ exist, but their exact forms are not important for the current discussion\footnote{Note that despite the word ``model'' used, we do not require that $1$-anyon is the zero-energy state of $\hat V_1^{\text{model}}$, but only that the $1$-anyon has lower energy than $\psi$-anyon with respect to this interaction. The same is true for the $\psi$-anyon and $\hat V_\psi^{\text{model}}$}.

Using this Hamiltonian, one can design an experiment to observe the effect of long-range entanglement as follows. Suppose, in a MR system with four quasiholes, $\sigma_1$,$\sigma_2$,$\sigma_3$,and $\sigma_4$, we pick one pair of quasiholes (say, $\sigma_1$ and $\sigma_2$) and move their positions to the same spatial point so that they fuse into a charge-$e/2$ quasihole. Afterward, we measure the braiding of $\sigma_3$ and $\sigma_4$. One can observe that the fusion outcome of $\sigma_1$ and $\sigma_2$ will affect the anyonic statistics of $\sigma_3$ and $\sigma_4$. Assume that the total system is in the even sector (described by a microscopic Pfaffian-type wavefunction with even number of electrons), then braiding $\sigma_3$ and $\sigma_4$ gives a phase of $\pi/2$ if $\lambda_1\ll\lambda_2$, but zero phase if $\lambda_1\gg\lambda_2$. This is because when the interaction favors the $\psi$-type quasihole, the ground state of the total Hamiltonian would contains two unfused $\sigma$-anyons, $\sigma_3$ and $\sigma_4$, and one $\psi$-anyon that is the result of fusing $\sigma_1$ and $\sigma_2$. In this case, because of parity conservation, $\sigma_3$ and $\sigma_4$ must also fuse into a $\psi$, and thus their braiding must gives $\pi/2$. The reverse is true if the interaction favors the $1$-type quasihole instead: the ground state consists of two unfused $\sigma$-anyons and one $1$-anyon, and braiding the two $\sigma$'s gives zero phase. In either case, the braiding phase becomes scalar and completely deterministic just by an introduction of a two-body interaction. This is in contrast to the case where there is no two-body interaction, or when the system contains only unfused $\sigma$-anyons: the ground state is always two-fold degenerate and braiding any two quasiholes results in either a scalar linear combination of $\pi/2$ and 0, or an $\text{SU}(2)$ rotation within the two-fold degenerate ground state due to its non-Abelian properties. 

A more complicated situation arises for six quasiholes. Suppose we start with $\sigma_i$, $i=1,...,6$. If all six quasiholes are far apart from each other, then the ground state is exactly four-fold degenerate. Now, if we fuse $\sigma_1$ and $\sigma_2$, then once again the species of anyons present in the ground states are governed by the effect of the specific two-body interaction. This time, the Hilbert space of $\sigma_3$, $\sigma_4$, $\sigma_5$, and $\sigma_6$ will always be two-fold degenerate, and in general non-Abelian braiding is expected in both outcomes of $\sigma_1$-$\sigma_2$ fusion. However, the two cases will have different braiding matrices: if the interaction favors the $1$-anyon ($\psi$-anyon) then $\sigma_1$ and $\sigma_2$ fuse to a $1$-anyon ($\psi$-anyon), so the braiding of the remaining four quasiholes will follow the odd (even) sector\cite{fukusumi2025operator},

The monodromy and braiding matrices of Moore-Read quasiholes in the even sector have been extensively studied, both analytically\cite{moore1991nonabelions,oshikawa2007topological,Bonderson2011,fukusumi2025operator} and numerically\cite{Tserkovnyak2003,wu2014braiding}, with the braiding matrices given explicitly for four quasiholes. On the contrary, the braiding matrices in the odd sector have not received the same attention, presumably due to the assumption that parity of a system is generally conserved by conservation of electron number. Here, we show that when considering a subsystem consisting of some but not all quasiholes, this assumption does not hold. The fusion of background quasiholes can result in odd-sector braiding behavior when the full system has even parity, and vice-versa. This happens if the following two conditions are satisfied: an odd number of pairs of background quasiholes are fused and electrons in the system interact via a two-body interaction that energetically favors the $\psi$-anyon (see Fig. \ref{fig:MR degeneracy splitting}). Neither of these conditions is unlikely in experiment as it is in general difficult to control the microscopic details of a sample to account for all quasiholes or to engineer a two-body interaction that perfectly balance the energies of the two fusion channels.

%The arguments above can be generalized into a system with $2n$ quasiholes for $n>2$. If we pick out a pair of quasiholes to fuse into a charge-$e/2$, then depending on the interaction the $2^{n-1}$ degenerate ground state will be split into two $2^{n-2}$ degenerate subspaces. Each subspace contains $2(n-1)$ $\sigma$-anyons and one charge-$e/2$ quasihole of either $1$ or $\psi$ type. The statistical properties of the space of the $2(n-1)$ $\sigma$'s would then depend on the type of anyon the charge-$e/2$ quasihole is, which is dependent on the microscopic reaction (see Fig.\ref{fig:MR degeneracy splitting}). 

%{\color{red} this is one of the most important part of the paper. add a specific discussion about six quasiholes, where non-abelian statistics is present. emphasize you are fusing one pair very far away, the braiding of the remaining 4 quasiholes, which is non-abelian, will be affected.}

\section{Energetics of MR quasiholes}
%{\color{red} transition from the previous and this section is not good}
Noting that the above discussion can be generalized to \emph{any} system hosting non-Abelian anyons, we now turn our focus to FQH physics. We explicitly demonstrate the effects of electron-electron interactions on non-Abelian anyon fusion channel selection and discuss the physical implications in braiding measurement. From a modern perspective, quasiholes of a given FQH phase can be viewed as the fundamental degree of freedom describing the physics within the nullspace of its model Hamiltonian\cite{yang2022anyons}. The FQH ground states and quasihole states make up a subspace of the LLL called conformal Hilbert space (CHS)\cite{wang2022analytic,yang2022anyons}. Every state in the CHS is the exact zero-energy state of some model Hamiltonian. When additional interactions are introduced, they manifest as effective self-energy of a quasihole and effective interactions between quasiholes\cite{trung2021fractionalization, yang2022anyons,xu2024}.  Understanding the physics of the FQH system in terms of quasiholes allows us to understand many phenomena observed in real experiments.

Here we apply this understanding to investigate the effect of two-body electronic interactions on the energetics of fusion channels. This allows us to predict their behaviors under any general two-body interaction (such as bare or screened Coulomb interactions) and with any anyon configuration. The advantages of our approach over the previous studies may enable  possibilities of designing braiding schemes that stabilize non-Abelian statistics.

\subsection{Effects of two-body pseudopotentials}
To start our discussion on the physics of MR anyons in realistic system, we consider their dynamics with respect to two-body interactions. Any general two-body interaction can be written in terms of the Haldane pseudo-potential\cite{haldane1983fractional}:
\begin{equation}
\label{Haldane pp}
\hat V^{2bdy}=\sum_{m}c_m\hat V_m^{2bdy}
\end{equation}
where $\hat V_m^{2bdy}$ is the $m$-th Haldane pseudopotential ($m=0,1,2,...$), which punishes a pair of electrons with relative angular momentum $m$, and $c_m$ are real coefficients. %Since electrons are fermions, $c_m=0$ for even number $m${\color{red} no they are non-zero, just don't have any physical effects}. 
We consider the self-energy, which is the energy required to create an anyon from the ground state. If the electronic interaction is the three-body model Hamiltonian, then this energy is zero for any number of anyons of any species. However it is non-zero if the electronic interaction is two-body. The self-energy of the 1-type and $\psi$-type anyons can be calculated numerically by calculating the variational energies of the wavefunctions given by Eq.(\ref{eq:MR 1qh}) and (\ref{eq:MR 1qh psi}), respectively. On the other hand, since it is impossible to create an isolated $\sigma$-anyon, its self-energy can be approximated as half of the variation energy of Eq.(\ref{eq:MR 2qh}) in the thermodynamic limit\cite{xu2024}.  

\begin{figure}
    \centering
    \includegraphics[width=1\linewidth]{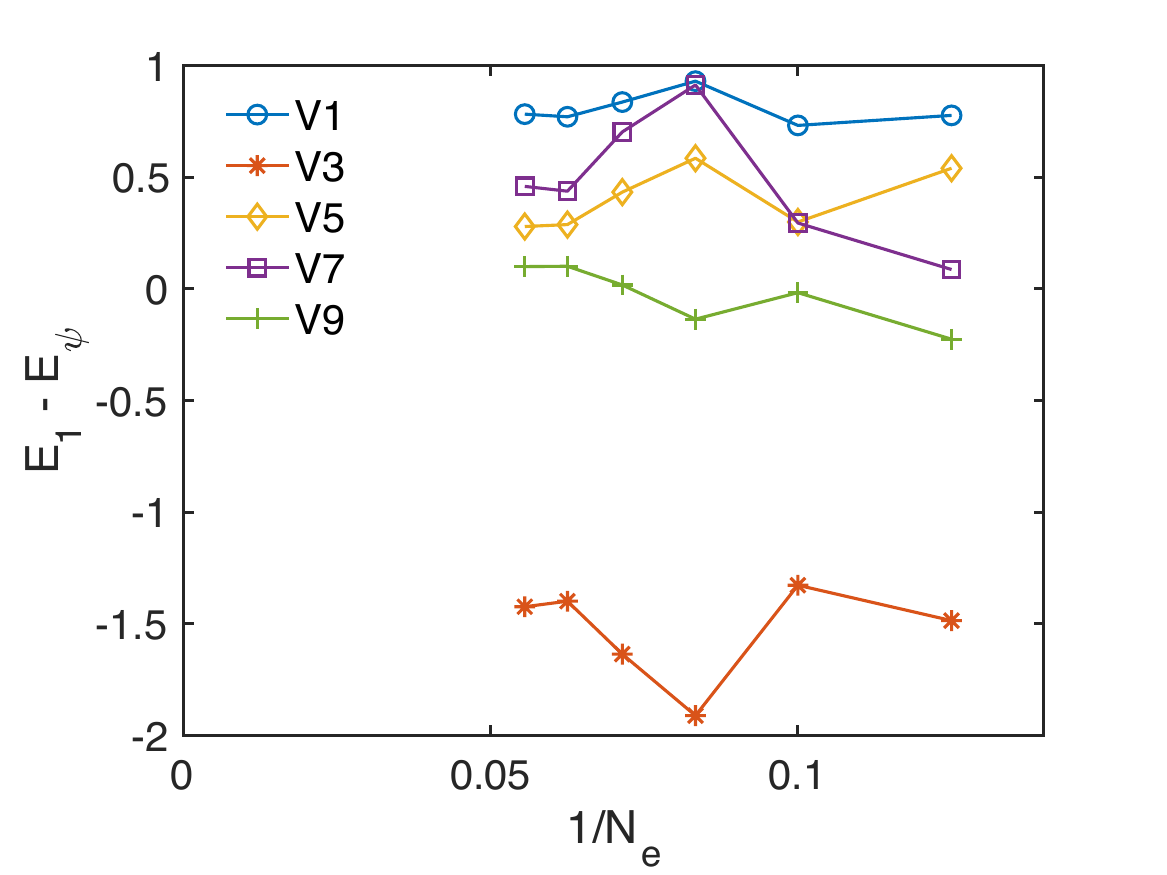}
    \caption{Difference in self-energy of the two types ($E_1-E_\psi$) with respect to $\hat V_{m}^{2bdy}$ for $m=1$ (blue circles), $m=3$ (red stars), $m=5$ (yellow diamons), $m=7$ (purple squares), and m=9 (green crosses). Maximum system size used in calculation has $N_e=18$ electrons. A positive energy difference implies that the $\psi$-type quasihole is energetically favorable compared to $1$-type, and vice versa\cite{rawdata}.}
    \label{fig_self-energy}
\end{figure}

%The self-energy of a single 1-type or $\psi$-type anyon can be computed numerically by evaluating the variational energies of the wavefunctions described in Eq.(11) and (12), respectively{\color{red} is this repetitive?}. As the system size increases, the interaction between a pair of 1-type or $\psi$-type anyons diminishes. In the thermodynamic limit, the interactions vanish due to the infinite separation between the two same-type anyons, leaving only their self-energies. 
The self-energy difference between the 1-type and $\psi$-type anyons, $E_1 - E_\psi$, is shown in Fig. \ref{fig_self-energy}, revealing which type of anyon is energetically favored. A positive (negative) difference indicates that the $\psi$-type (1-type) is energetically favored. Remarkably, the energy difference is relatively consistent across different system sizes, allowing us to make qualitative conclusion about their behavior in the thermodynamic limit despite the computational limitation. We see that the pseudopotential $\hat V_3^{2bdy}$ strongly prefers the 1-type anyon, while the other pseudopotential $V_m^{2bdy}$, $m\neq3$, up to $m=9$ prefers the $\psi$ type. 

Typically, the coefficients $c_m$ in Eq.(\ref{Haldane pp}) decreases for increasing $m$\cite{haldane1983fractional,cage2012quantum} for realistic interactions. We also see a trend in Fig. \ref{fig_self-energy} that the self-energy difference between the two anyon species decreases with increasing $m$ (for $\hat V_9^{2bdy}$ the difference is almost zero). It is thus sufficient to consider only the effects of the first few pseudopotentials. Notably, only $\hat V_3^{2bdy}$ favors the $1$-type anyons, while other pseudopotentials favor the $\psi$-anyons (with $\hat V_1^{2bdy}$ favoring it most strongly). Understanding the effect of each individual potential is useful for investigating the effects of different tuning parameters, as we present below. Additionally, as several works have suggested the possibility of realizing individual pseudopotentials in experiments\cite{seiringer2020emergence,yang2024emergent}, the strong opposing effects of $\hat V_1^{2bdy}$ and $\hat V_3^{2bdy}$ may become the essential tools for engineering anyon dynamics in the MR system.

\begin{figure*}
\centering
\includegraphics[width=\linewidth]{fig6.pdf}
\caption{(a) Schematic diagram showing the procedure carried out in numerics. In a system with four MR quasiholes, two quasiholes are each trapped by two potential pins and moved along a circular exchange path on the northern hemisphere, while the other two are fused into a charge-$e/2$ quasihole and fixed by a strong potential pin at the south pole. (b-g) The spectrum of Eq.(\ref{braiding Hamiltonian}) plotted against parameter $\theta_0$ for MR systems with: (b) 10 electrons, $\lambda_1=0.1$ (c) 10 electrons, $\lambda_1=1.0$, (d) 10 electrons, $\lambda_1=10.0$, (e) 11 electrons, $\lambda_1=0.1$, (f) 11 electrons, $\lambda_1=1.0$, (g) 11 electrons, $\lambda_1=10.0$\cite{rawdata}.}
\label{MR spectrum}
\end{figure*}

\subsection{Effect of quasihole energetics on braiding}
\begin{figure*}
    \centering
    \includegraphics[width=\linewidth]{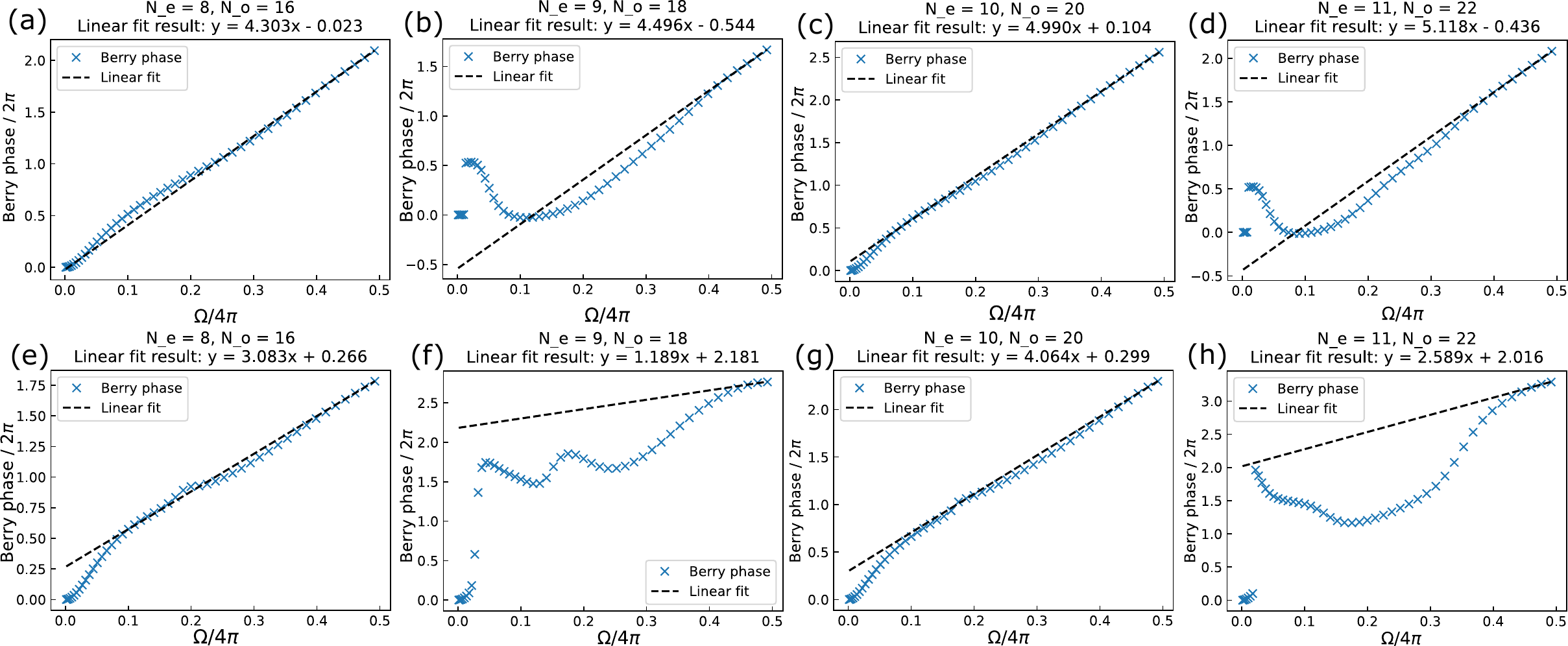}
    \caption{Berry phase as a function of solid angle $\Omega=2\pi(1-\cos\theta_0)$ in the braiding scheme using Eq. \ref{braiding Hamiltonian} described in the main text. A linear fit is performed on a few right-most data points ($\Omega\sim0.5$), from which the exchange statistics in multiple of $2\pi$ can be taken as the $y$-intercept (shown in the title of each respective plot) (a) 8 electrons, $\lambda_1=0$ (b) 9 electrons, $\lambda_1=0$ (c) 10 electrons, $\lambda_1=0$ (d) 11 electrons, $\lambda_1=0$ (e) 8 electrons, $\lambda_1>0$ (f) 9 electrons, $\lambda_1>0$ (g) 10 electrons, $\lambda_1>0$ (h) 11 electrons, $\lambda_1>0$\cite{rawdata}.}
    \label{fig:MR numerics}
\end{figure*}

Having determined the effects of different pseudopotential terms on favoring a certain type of fused anyon, we can proceed to verify the effect of fusion channel selection on measured braiding statistics by numerical simulation. Here, we employ exact diagonalization (ED) on a system with four quasiholes. Once again, we work on the spherical geometry where position is parametrized by an azimuthal angle $\theta$ and a polar angle $\phi$. We consider a potential pin configuration consisting of three pins, each with a profile described by the Dirac delta function $\delta^2(\theta-\theta_0,\phi-\phi_0)$ centered at $(\phi_0,\theta_0)$. One pin of strength $g_1$ fixed at the south pole, and two pins of strength $g_2$ located at positions $(\theta,\phi)=(\theta_0,\phi_0(t))$ and $(\theta_0,\phi_0(t)+\pi)$ (see Fig. \ref{fig:MR numerics}a). The polar position $\phi_0(t)$ is then varied between 0 and $\pi$ such that two pins are exchanged along a circle of latitude, which is mathematically equivalent to a rotation by $\pi$ around the $z$-axis. Thus, $\hat L_z$ of the ground state is proportional to the total Berry phase obtained. The exchange phase can then be extracted from this Berry phase in the same manner described in Ref.\cite{trung2023spin}. The Gaussian profile is chosen for each potential pin because its matrix element in the Landau orbital basis can be computed analytically. All in all, the Hamiltonian used in ED is\cite{seesup}
\begin{align}
    \hat H(\theta_0) &= \lambda_0\hat V_3^{3bdy}+\lambda_1\hat V_1^{2bdy} +\hat H_{\text{pins}}(\theta_0)\label{braiding Hamiltonian}\\
    \hat H_{\text{pins}}(\theta_0) &= g_1 \hat V_{g}(\pi,0)+g_2\left[\hat V_{g}(\theta_0,0)+\hat V_{g}(\theta_0,\pi)\right]\\
    \hat V_{g}(\theta_0,\phi_0)&=\int d\Omega\hat \rho(\theta,\phi)\delta^2(\theta-\theta_0,\phi-\phi_0) \label{one pin 1}
\end{align}
Here $\hat\rho(\theta,\phi)$ is the density operator in real space. In our numerics we set $g_2=2g_1$ so that the pin at the south pole always traps a charge-$e/2$ quasihole, while each of the other two pins traps a charge-$e/4$ quasihole. 

In our numerics, we take the limit $\lambda_0\to\infty$ so that all  low-lying states reside within the MR CHS spanned by the Jack polynomials. Even then, there are two possible tuning parameters: $\lambda_1$, which is the strength of $\hat V_1^{2bdy}$, and $g_1$, which is the strength of the potential pin. We fix $g_1=1$ and investigate the effect of tuning $\lambda_1$ on the spectrum. Based on the previous analysis, we know that $\hat H_{\text{pins}}$ slightly favours the $1$-anyon energetically while $\hat V_1$ favours the $\psi$-anyon, and thus we expect a crossing between the lowest two energy levels when the charge-$e/2$ anyon pinned at the south pole transmutes from $1$-type to $\psi$-type. However it is not so clear in numerics, as seen in the spectrum in Figs. \ref{MR spectrum}b-g. This is because the spectrum is complicated by the presence of the two charge-$e/4$, which interacts at finite separation. The behavior in the even sector ($N_e=10$ electrons, Figs. \ref{MR spectrum}b-d) is relatively simple: at $\theta_0=0$, the ground state is either two $1$-anyons, one at each pole, for small $\lambda_1$, or two $\psi$-anyons, also one at each pole, for large $\lambda_1$. As $\theta_0$ is increased (at fixed $\lambda_1$), the charge-$e/2$ quasihole pinned at the south pole remains a fixed type, and the interaction between the two charge-$e/4$ results in the fluctuation in energy of the ground state (the quantitative behavior of this effective anyon interaction is known from Ref. \cite{xu2024}). 

On the other hand, the odd sector ($N_e=11$ electrons, Fig. \ref{MR spectrum}e-g is more complicated since at $\theta_0$=0 the ground state is two-fold degenerate: in the system there is one $1$-anyon and one $\psi$-anyon, the former of which may be pinned at either the north pole or the south pole, and the latter will be pinned at the opposite pole. Because of this, when $\theta_0$ are increased, the charge-$e/2$ quasihole at the south pole may spontaneously transmute between the two anyon types, resulting in level crossings between the two lowest energies. However, since these level crossings are a results of balancing between interaction between the two charge-$e/4$ anyons and the self-energy of the charge-$e/2$ anyon, they are not expected to occur at large $\theta_0$, where the charge-$e/4$ quasiholes are separated enough from each other such that they are non-interacting. Indeed, in Figs. \ref{MR spectrum}e-g we see that the crossings only occur at relatively small values of $\theta_0$ (less than 0.5 radian).

The Berry phase resulting from this braiding scheme is shown in Fig. \ref{fig:MR numerics} for systems with $N_e=8,9,10$, and $11$ electrons. We first consider the even sector ($N_e=8$ and $10$) and $\lambda_1$=0: the ground state in a system with four MR quasiholes is not an exactly two-fold degenerate, as one may expect. This is because the south-pole pin slightly favor the $1$-anyon due to its lower local electron density\footnote{This is a unique feature of the Dirac delta potential profile we used. In real experiment, if the one-body potential has a larger profile (e.g. a size scale of several magnetic length) the energy gap between two types of quasiholes could be smaller and the non-Abelian manifold is preserved.}. As a result, the other two charge-$e/4$ quasiholes must also fuse into a $1$-type so that the total number of $1$-anyon is even. Since the ground state is now Abelian (it is non-degenerate), the braiding phase of two quasiholes can be related to their topological spins, as well as the topological spin of their fusion product. In this case when two MR quasiholes fuse into a $1$-anyon, one observe a trivial braiding phase between them.

Now we turn on $\hat V_1^{2bdy}$ by setting $\lambda_1>0$. Since the $\psi$-type has lower variational energy with respect to $\hat V_1^{2bdy}$, as we increase $\lambda_1$ there will be level-crossing point beyond which the ground state consists of a $\psi$-type quasihole trapped at the south pole. Now, the free quasiholes must fuse into $\psi$ type so that the total number of $1$-anyon is zero, an even number. Thus, exchanging the free quasiholes result in an exchange phase of $\pi/2$. This change in braiding phase is reflected in the qualitative difference between Figs. \ref{fig:MR numerics}a and e for 8 electrons and between Figs. \ref{fig:MR numerics}c and g for 10 electrons. (In this numerics, we choose $\lambda_1=1$ for the even sector so that the spectrum is qualitatively different from the case when $\lambda_1=0$ , see Fig. \ref{MR spectrum}c.) Note that the braiding phase extracted by this method is not exactly quantized at either $0$ or $\pi/2$ due to finite size effect\cite{trung2023spin,seesup}. However, one can see the qualitative difference in the braiding phase $\gamma_{br}$ between $\gamma_{br}\sim0$ at $\lambda_1=0$ and $\lambda_{br}\sim\pi/2$ at $\lambda_1>0$

We also note that the qualitative effect of $\hat V_1^{2bdy}$ can also be seen in the odd sector ($N_e=7,9,11,...$), though the quantitative result depends on the complex interaction among all three types of anyons and remains to be studied further. In this case, the braiding phase is $\pi/2$ if $\lambda_1=0$ and $0$ if $\lambda_1>0$, the opposite of the results for the even sector. The numerical results, shown in Fig. \ref{fig:MR numerics}b and f for 9 electrons and Fig. \ref{fig:MR numerics}d and h for 11 electrons, are less stable compared to their even-sector counterparts, as evidence in the large jump in the Berry phase data points. These jumps are a direct consequence of the level crossings between the lowest two energy states, as discussed above and seen in Fig. \ref{MR spectrum}e-g. This complicated anyon dynamics could play a crucial role in designing a suitable braiding scheme, which will be explored further in future work. However, we emphasize that anyon interaction at finite seperation \emph{not} affect our discussion about the entanglement between fused background anyons and measured braiding statistics, as we can see the qualitative difference between the braiding phases in Figs. \ref{fig:MR numerics}b and e (close to integer times $\pi$) and those in Figs. \ref{fig:MR numerics}c and d (close to half integer times $\pi$). In the thermodynamic limits, these quantities should converge to their expected values. This result shows an interesting ``long-range entanglement" behavior where what happens to a pair of anyon can affect others that are infinitely far away in a physically measurable manner.

%As can be seen in additional figures in the supplementary, this numerical result persists as the system size is increased. In particular we expect the behaviour of the exchange statistics to hold in the thermodynamic limit, in which the south pole is infinitely far away from the equator, when the exchange takes place. This is expected because our argument in the previous section about fusion channel selection makes no mention of the spatial separations between the anyons. In reality, this shows an interesting ``long-range entanglement" behavior where what happens to a pair of anyon can affect others that are infinitely far away in a physically measurable manner. 

An important message here is that even though theoretically one expect the four-quasihole Hilbert space to be doubly degenerate, which gives rise to non-Abelian statistics, this degeneracy can be split if the following two conditions are fulfilled: (i) the number of pins is less than the number of quasiholes (i.e. fewer than four pins), and (ii) the two-body interaction energetically favors one type of charge-$e/2$ quasihole over the other. Therefore, an experiment aimed to observe the non-Abelian statistics must either account for all charge-$e/4$ in the systems (ensure that they are all well-separated from each other), or fine tune the two-body interaction such that both types of charge-$e/2$ quasiholes have the same creation energy. The first method requires us to apply the same number of potential pins as the number of quasiholes in the system, which is extremely challenging since it is in general very difficult to determine the exact number of quasiholes in a given experimental system. The second method, however, is possible based on our analysis in the previous section.

We stress that while the effect of ground-state degeneracy on statistics of MR quasiholes have been discussed before, past literature focused on the effect of fusion channel selection of the same pair of $\sigma$-quasiholes being braided\cite{Prodan2009,baraban2009numerical}. The fusion in that context can then be thought of as a ``virtual'' process which helps to determine the braiding phase via the spin-statistics relation for anyons\cite{trung2023spin,nardin2023spin}. Here, we show that the fusion of one pair of quasiholes in real space (and real time) can affect the braiding of other quasiholes in the system. This phenomenon may present many challenges in experiments, as the configuration of background quasiholes may be heavily influenced by impurities in the sample, which is beyond control.

\subsection{Tuning the non-Abelian gap in disordered systems}

We now turn attention to the ``non-Abelian gap'', which we define as the energy gap between different orthogonal states within the non-Abelian degeneracy with respect to some two-body electron interaction. Non-Abelian statistics can only be realized when this gap is much smaller than the thermal energy (which in turn cannot be so large that topological phase is completely destroyed). This motivates our discussion on tuning the electron interaction to realize this limit. To simplify the discussion, we consider in this section the system with four MR quasiholes, where a two-fold degeneracy is expected when the positions of all quasiholes are fixed by the one-body potential. However, if a charge-$e/2$ is present anywhere in the system, then this degeneracy is split by the effect of two-body interaction as discussed in the section above. The toy-model Hamiltonian used in this section is
\begin{equation}
\label{Hamiltonian with pins}
\hat H = \lambda_0 \hat V^{3bdy}+\hat V^{2bdy} + v_0\hat V_{\text{pins}}
\end{equation}
Once again, we take the limit $\lambda_0\to\infty$ so that the low-lying energy eigenstates reside completely within the MR CHS. Let us consider the setup where the Hall bar contains four $\sigma$-type quasiholes, and $\hat V_{\text{pins}}$ consists of two potential pins, each with a profile of a Dirac delta function and each very well separated from the others. On the sphere, this is implemented by placing one potential pin at the north pole and one at the south pole:
\begin{equation}
\label{two pins}
\hat V_{\text{pins}} = \hat V_g(0,0) + \hat V_g(\pi,0)
\end{equation}
where $\hat V_g(\theta_0,\phi_0)$ is the potential pin centered at $\theta=\theta_0$ and $\phi=\phi_0$ defined in Eq.(\ref{one pin 1}). The factor $v_0$ in Eq.(\ref{Hamiltonian with pins}) is a tuning parameter that signifies the strength of the pinning potential. As we shall see, the interplay between one- and two-body potentials play a crucial role in minimizing the non-Abelian gap.

%\subsection{Interplay between one-body and two-body potentials}
Let us start by considering a simple example, where the two-body interaction consists only of the first two pseudopotential. Within the MR CHS the Hamiltonian is
\begin{equation}
\label{v1v3}
\hat H = (1-\lambda) \hat V_1^{2bdy} + \lambda \hat V_3^{2bdy} + v_0\hat V_{\text{pins}}
\end{equation}
where $\lambda$ is a tuning parameter such that the interaction interpolates between $\hat V_1^{2bdy}$ and $\hat V_3^{2bdy}$ for $\lambda$ going from 0 to 1. 

\begin{figure}
\begin{center}
\includegraphics[width=\linewidth]{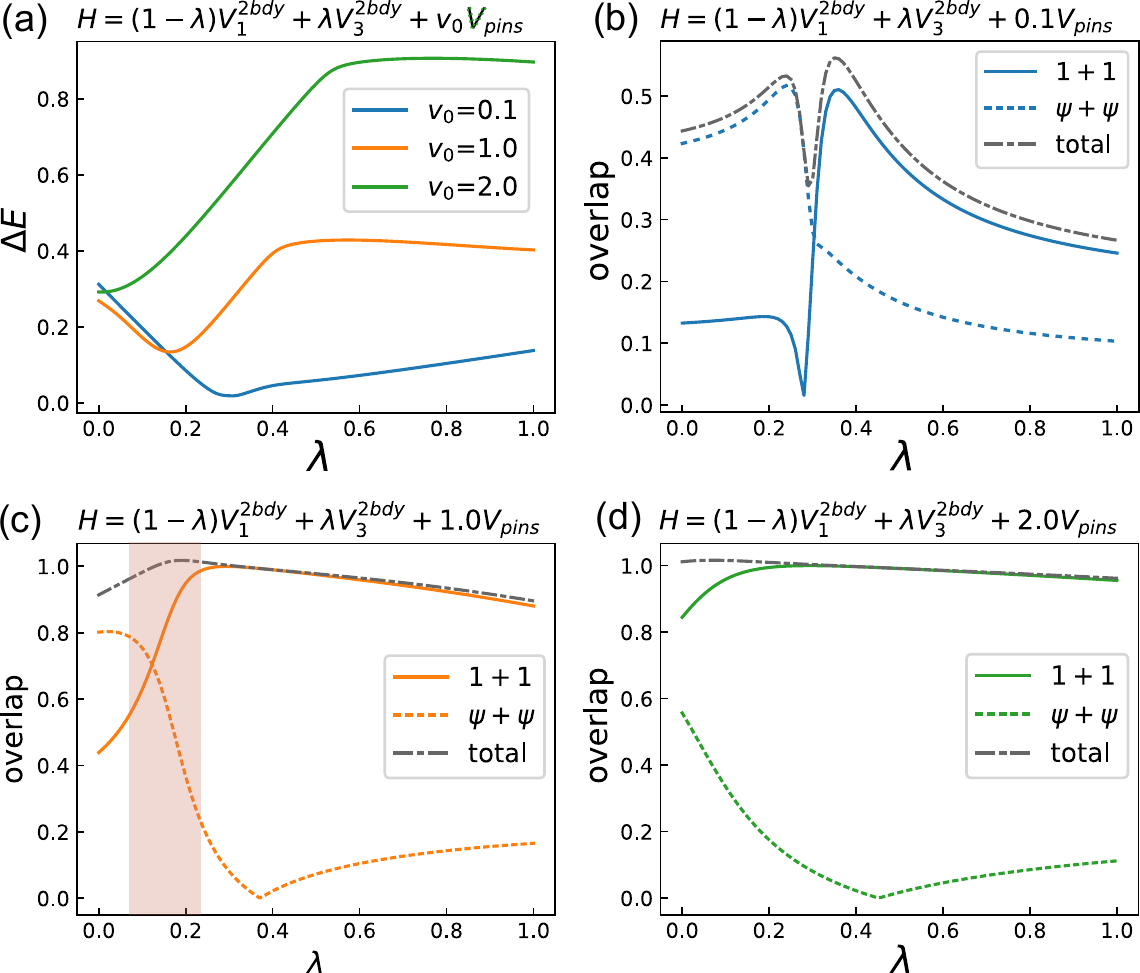}
\caption{(a) Energy gap between the lowest two eigenstates of the Hamiltonian given by Eq.(\ref{v1v3}) plotted against $\lambda$ for $v_0=0.1$ (blue), $v_0=1.0$ (orange), and $v_0=2.0$. The total strength of two-body interactions is normalized to unity, so $v_0$ represents relative strength of the one-body potential compared to two-body interaction. Numerical data were otained by exact diagonalization in the $L_z=0$ sector for a MR system consisting of 12 electrons and 24 orbitals (b)-(d) The overlaps of the ground state of Eq.(\ref{v1v3}) with the model state of two $1$-type anyons (solid lines) and with the model state of two $\psi$-type anyons (dashed lines), and the total overlap with these two states (dot-dashed lines) plotted against $\lambda$ for (b) $v_0=0.1$ (c) $v_0=0.5$ (d) $v_0=2.0$. Boxed overlays denotes a parameter region ideal for detecting non-Abelian statistics\cite{rawdata}.} 
\label{fig:v1v3 gap}
\end{center}
\end{figure}

All in all, in this toy model we have two tuning parameters: the relative strength between $\hat V_1^{2bdy}$ and $\hat V_3^{2bdy}$, and the relative strength of the pinning potential compared to the total strength of the two-body interaction. Here we focus on the case where only charge-$e/2$ anyons are present in the system, and it is useful to recall the effect of each of the terms in the Hamiltonian. Due to the different density profiles of the two anyon types (see Fig. \ref{fig:MR quasiholes}b), the pinning potential energetically favors the two types of anyons differently. Specifically, the $1$-anyon has lower electron density at the center, so a pinning potential with an extremely thin profile, such as those used in this study in Eq.(\ref{two pins}), energetically favors the $1$-anyon over the $\psi$-anyon. As for the two-body interactions, $\hat V_1^{2bdy}$ prefers the $\psi$-anyon while $\hat V_3^{2bdy}$ prefers the $1$-anyon, as discussed in the previous section. The effects and physical meanings of these terms are also summarized in Table \ref{summary table}.
\begin{table*}
\begin{tabular}{|c|c|c|}
\hline
Potential & Physical meaning & Effect\\
\hline
$\hat V_1^{2bdy}$ & Extremely short-ranged (hard-core) electron-electron interaction & prefers $\psi$-type quasihole\\
$\hat V_3^{2bdy}$ & Longer-ranged (``hollow-core'') electron-electron interaction & prefers $1$-type quasihole\\
$\hat V_{\text{pins}}$ & Pinning potential (e.g. from an STM tip) & prefers $1$-type quasihole\\
\hline
\end{tabular}
\caption{Summary of the effects of different terms in Eq. (\ref{v1v3}) and how they relate to tunable experimental parameters.}
\label{summary table}
\end{table*}

We can see the effect of the interplay between these three potentials in Fig. \ref{fig:v1v3 gap}a, which shows the gap between the lowest two eigen-energies of Eq. (\ref{v1v3}) for a system consisting of 12 electrons and 24 orbitals (similar trends are obtained for systems with 8 and 10 electrons, as shown in the supplementary materials). With this system there are four MR quasiholes, which pair-wise form one charge $e/2$ at the north pole and another at the south pole. In this choice of potential pins configuration the system exhibit rotational symmetry about the $z$-axis, which allows us to further reduce the the necessary Hilbert space used in numerics from the full MR CHS to only the $L_z=0$ sector. 

In Fig. \ref{fig:v1v3 gap}a one can observe the combined effect of everything we have learned about quasihole energetics. If the potential pin strength $v_0$ is too large, such that it dominates over the two-body interaction, then the ground state is always the ``$1+1$'' state (given by Eq.(\ref{two 1-types})), which consists of two $1$-anyons. However, if the trapping pins are not so trong and the two-body interaction is dominated by $\hat V_1^{2bdy}$, it is possible that the ground state is ``$\psi+\psi$'' (given by Eq.(\ref{two psi-types})) instead. In this weak pinning potential limit, if one starts adding in $\hat V_3^{2bdy}$ by tuning $\lambda$ in Eq. (\ref{v1v3}), the Hamiltonian will start to favor the ``$1+1$'' state again. In the limit $\lambda\to1$ the Hamiltonian completely favors the ``$1+1$'' state agains since both $\hat V_3^{2bdy}$ and $\hat V_{\text{pins}}$ favor it. Thus, there must be a point in the middle of the two limits where a crossing between the two states happen. We can see this crossing point in Fig. \ref{fig:v1v3 gap}b-c (it is also non-existent in Fig. \ref{fig:v1v3 gap}d in the strong potential pin limit). It is noteworthy that the gap does not always fully close at the crossing point, which could be a band repulsion effect coming from anyon-anyon interaction in a finite system. However, at the crossing point the gap reaches its minimum, which can be either vanishing in the thermodynamic limit, or potentially overcome by thermal energy in real experiments. Most notably the total overlap with the two model states (defined by the root sum squared of the two overlaps) is very close to unity near the crossing point. This indicates that the two lowest, near-degenerate states are fully described by the model states for the $1$-type and $\psi$-type anyon. Potentially, this means that this parameter range (denoted by an boxed overlay in Fig. \ref{fig:v1v3 gap}c) is ideal for carrying out non-Abelian braiding. However, to be able to confidently make this statement requires further analysis into scenarios also involving charge-$e/4$ anyons, which will be presented elsewhere.

While the interaction in Eq. (\ref{v1v3}) is purely artificial, our analysis serves as a proof-of-concept that in principal it is possible to make the non-Abelian gap approach zero by tuning the two-body interaction. Our results also suggest a range for the ratio between the amounts of $\hat V_1^{2bdy}$ and $\hat V_3^{2bdy}$ in order to achieve this. This can serve as a rough estimation for the target range of electron-electron interaction for stabilizing non-Abelian properties in experiments. In real experiments, this ratio can be tuned by tuning the range of the effective electron-electron interactions. Roughly speaking, a longer-ranged interaction implies higher $\hat V_3^{2bdy}$-to-$\hat V_1^{2bdy}$ ratio. This is possible by one of the two methods: either tuning the finite sample thickness, or tuning the strength of the screening of the Coulomb interaction. In the supplementary material, we provide a brief quantitative discussion on the effects of these two parameters, using the the Zhang-Das Sarma (ZDS) interaction to model the effect of finite sample thickness\cite{zhang1986excitation} and the Yukawa potential to model Coulomb screening.

\section{Conclusion}
Our study points out an interesting phenomenon where the parity conservation in a MR system manifests as a long-range entanglement between anyons. It is argued and shown numerically that the fusion of one pair of anyons somewhere in the system can affect the statistics of others in the systems, even if all anyons are infinitely far apart from each other. Note that this is a topological effect that is not due to any geometrical effect such as deformation or long-tail overlapping, as evident from finite-size scaling. Our finding implies that in experiments involving non-Abelian anyons, it is not enough to only look at anyons in a sub-system (e.g. only within the area between two quantum point contacts in interferometry experiments\cite{Nakamura2020, willett2023interference}), but one has to account for all anyons on the Hall bar in order to make accurate predictions about experiment outcomes. Whenever that is not possible, a simple workaround might be to use the proposed tuning parameters to ensure exact degeneracy of the two fusion channels.

Our analysis of anyon energetics also provides a tool for possibly fine-tuning the ground state degeneracy in MR systems, which is crucial for detecting signature non-Abelian statistics. This involves balancing the variational energies of the two fusion channels with respect to different two-body pseudopotentials. By tuning the range of the effective electron-electron interaction, such as by screening the Coulomb interaction or increasing sample thickness, and tuning the strength of the pinning potentials, one can achieve a regime with an ideal non-Abelian degeneracy. Our argument relies on treating the quasihole excitations of the MR state as the fundamental degrees of freedom describing the physics of low-lying states. In this picture two-body electron interactions becomes the quasihole self-energies. While we focus here on the case where all quasiholes are well-separated, our understanding can also be generalized to the case of finite separation, since it has been shown that in that case, electron-electron interaction manifests as
effective interactions between quasiholes\cite{trung2021fractionalization,xu2024}. Generalization to other non-Abelian states is also possible with similar analysis.

This study motivates further analysis of braiding properties within the odd sectors. Furthermore, in finite-time braiding processes (as opposed to braiding over an infinitely long time period, ensuring adiabaticity), the splitting energy gap of the two fusion channels potentially contribute to the outcome of the braiding. Quantitative analysis of this contribution, as well as calculating the Berry matrices in different braiding schemes remain an open question for future studies.

\section*{Acknowledgement}
HQT would like to acknowledge Ji Guangyue and Nilanjan Roy for meaningful discussions and Truman Yu Ng for helpful comments on an earlier draft of this manuscript. This work is supported by the National Research Foundation, Singapore under the NRF
Fellowship Award (NRF-NRFF12-2020-005), Singapore
Ministry of Education (MOE) Academic Research Fund
Tier 3 Grant (No. MOE-MOET32023-0003) “Quantum
Geometric Advantage”, and Singapore Ministry of Education (MOE) Academic Research Fund Tier 2 Grant
(No. MOE-T2EP50124-0017).
%\nocite{*}
\bibliography{reb}

%%%%%%%
%
% 			SUPPLEMENTARY GOES HERE
%
%%%%%%%
\clearpage

\onecolumngrid

\renewcommand{\thesection}{S\arabic{section}}
\renewcommand{\thefigure}{S\arabic{figure}}
\renewcommand{\theequation}{S\arabic{equation}}
\renewcommand{\thepage}{S\arabic{page}}
\setcounter{figure}{0}
\setcounter{page}{0}

\begin{center}
{\large \textbf{Supplementary Material for ``Long-range Entanglement and Role of Realistic Interaction in Braiding of Non-Abelian Quasiholes in Fractional Quantum Hall Phases''}}

\vspace{1cm}

\noindent\mbox{%
    \parbox{0.8\textwidth}{%
        \indent 
We first illustrate parity conservation from the Jack polynomial root configurations for six and more quasiholes. Next, we restate the main message of the paper in a more visual language, borrowing the tree diagrams used in conformal field theory\cite{bonderson2008interferometry,Bonderson2011}. The subsequent sections contain details on the numerical calculations presented in the main text, including evaluation of creation energy as well as details of the braiding process. Additional results for numerical simulations with realistic electron-electron interaction and impurities are also shown.
    }%
}
\end{center}

\section{More on parity conservation}
\subsection{Jack polynomials for the MR states}
To make the paper more self-contained, we start by recalling the basic properties of Jack polynomials and introduce relevant terminologies. Let us start by considering the following binary string
\begin{equation}
\label{MR root}
11001100110011001100110011
\end{equation}
This binary string can be treated as a representation of electrons occupying the Landau orbital. Each digit corresponds to an orbital, which on the spherical geometry used throughout this paper corresponds to angular momentum quantum numbers $L_z=S,S-1,S-2,...,-S+2,-S+1,-S$ where $S$ is an integer or half-integer. A digit $1$ means that the orbital is occupied by an electron while a digit $0$ means it is empty. As per our convention, the left-most orbital corresponds to the north pole ($L_z=S$) and the right-most orbital corresponds to the south pole ($L_z=-S$). 

One can check that Eq.(\ref{MR root}) satisfies the $(2,4)$-admissibility rule: there exists no more than two electrons within four consecutive orbitals (i.e. no more than two $1$'s within any four conseuctive digits). Any $(2,4)$-admissible configuration can serve as a ``root configuration'' for a Jack polynomial that normalizes to the MR state. Given a root configuration and an accompanying rational parameter $\alpha$, which equals to $-2$ for MR states, the monomial coefficients of a Jack polynomial can be computed recursively\cite{bernevig2008model, bernevig2008generalized}. 

Let us denote $N_o$ the number of orbitals (number of digits) and $N_e$ the number of electrons (number of $1$'s). We can see that Eq.(\ref{MR root}) satisfies the \emph{commensurability condition} $N_o=2N_e-2$. This is the unique root configuration that minimizes the ratio $N_e/N_o$ while satisfying the $(2,4)$-admissible rule --- there exists no $(2,4)$-admissible ocnfiguration with $N_o<2N_e-2$. Jack polynomials with such root configurations give the ground state of the MR state. Quasihole states are obtained by Jacks whose root configurations have $N_o>2N_e-2$. If a positive integer $n$ is the number of added orbitals to the ground state, i.e. $N_o=2N_e-2+n$, then $2n$ quasiholes are created.

\subsection{Parity conservation seen from circles and crosses}
Let us consider all the possible roots with $N_e$ electrons and $N_e=2N_e-2+n$ orbitals (again $n$ is some positive integer). Unlike the ground state, whose $(2,4)$-admissibility rule and commensurability condition can only be simultaneously satisfied when $N_e$ is even, quasihole states exist for odd $N_e$ as well. The list of all possible admissible configurations can be obtained by starting with a state with the lowest $L_z$:
\begin{align}
\underbrace{000..0}_{n}1100110011...110011&&\text{(even $N_e$)}\\
\underbrace{00..0}_{n-1}1001100110011...110011&&\text{(odd $N_e$)}
\end{align}
where the ellipses ``$...$'' denotes repeating patterns of $1100$. Afterward, all other admissible root can be generated recursively by picking a digit $1$ and move it one unit to the left, provided both that the digit to its left is $0$ and that the resulting configuration does not violate the admissible rule. Following this procedure, the first few terms go like
\begin{align}
\underbrace{000..0}_{n}1100110011...110011\label{even1}\\
\underbrace{00..0}_{n-1}10100110011...110011\label{even2}\\
\underbrace{00..0}_{n-1}11000110011...110011\label{even3}\\
\underbrace{0..0}_{n-2}100100110011...110011\label{even4}\\
\underbrace{0..0}_{n-2}100101010011...110011\label{even5}\\
\underbrace{0..0}_{n-2}101000110011...110011\label{even6}
\end{align}
for even $N_e$, and 
\begin{align}
\underbrace{00..0}_{n-1}1001100110011...110011\\
\underbrace{00..0}_{n-1}1010100110011...110011\\
\underbrace{00..0}_{n-1}1010101010011...110011\\
\underbrace{00..0}_{n-1}1100100110011...110011\\
\underbrace{0..0}_{n-2}10010100110011...110011
\end{align}
for odd $N_e$.

The parity conservation discussed in the main text is based on an empirical observation about the quasihole positions in root configurations of the Jacks. In the root configuration, a quasihole (charge $e/4$) occurs whenever four consecutive orbitals contain fewer than two electrons. Using this rule, one can mark down the positions of the quasiholes on the root configuration. Take Eqs.(\ref{even1}) to (\ref{even4}) with $n=3$ as an example:

\begin{align}
\subringleft{\blankchar}\quadsubring{0}0\subringleft{0}1100110011...110011\label{even1marked}\\
\subringleft{\blankchar}\doublesubringleft{0}\doublesubring{0}10\subringright{1}00110011...110011\label{even2marked}\\
\subringleft{\blankchar}\doublesubringleft{0}\subringleft{0}110\doublesubring{0}0110011...110011\label{even3marked}\\
\doublesubring{\blankchar}0\doublesubx{1}00\doublesubx{1}00110011...110011\label{even4marked}
\end{align}
A quasihole is marked at the center of each group of four digits with either one symbol (if the group contains one $1$) or two symbols (if the group contains two $1$'s). The choice of symbol can either be circles (`$\circ$'') or crosses (``$\times$'') based on the following rule: we first divide $2n$ quasiholes into $n$ pairs, going from left to write, then mark each pair with circles if there are an even number of $1$'s between them, or crosses if there are an odd number of $1$'s between them.
%\section{Internal structure of different anyon species}

Following our labeling rule it is easy to observe that the number of crosses must be an even number, say $2p$ where $p$ is some non-negative integer. With extensive inspection we can make a stronger observation that $p$ must be even if $N_e$ is even, and odd if $N_e$ is odd (for example, $p=0$ or $2$ in Eq.(\ref{even1marked})-(\ref{even4marked})). We claim that this is an empirical observation of parity conservation as two quasiholes marked by circles always fuse into a $1$-anyon and two quasiholes marked by crosses always fuse into a $\psi$-anyon. A few comments are in order regarding this observation:
\begin{itemize}
\item Our claim that two circles fuse into $1$ and two crosses fuse into $\psi$ is based on the observation of the root configuration for $n=1$. In that case, the root configuration of the lowest-$L_z$ states are:
\begin{align}
\doublesubring{\blankchar}01100110011...110011\\
\doublesubx{1}001100110011...110011
\end{align}
With only two quasiholes created, only a maximum one $\psi$-anyon may be created. Thus, the fusion rule $\psi\times\psi=1$ cannot occur (since it requires at least two $\psi$-anyons) and the fusion channel of two $\sigma$-anyons are fixed. The even sector (even $N_e$) contains no $\psi$ and the odd sector (odd $N_e$) contains exactly one $\psi$. By inspecting all possible root configurations, one can observe that the roots for even $N_e$ always contains two circles and the roots for odd $N_e$ always contains two crosses. Thus this support our argument that two circles fuse into a $1$ and two crosses fuse into a $\psi$.

\item In the rule for marking circles and crosses, we divide the quasiholes into pairs based on their order of appearance from left to right. \emph{A priori}, this is a rather arbitrary choice as in general a fusion channel exists for any pair of quasiholes. However, one can further check that no matter how $2n$ quasiholes are divided into $n$ pairs, as long as we mark each pair as two circles and two crosses based on whether the number of $1$'s between them is even or odd, respectively, the resulting numbers of circles and crosses are the same. Thus, the exact marking scheme does not affect our parity conservation, which only counts the total number of crosses regardless of where they occur in the root configuration. In other words, parity is well-defined in our ``circles-and-crosses'' description.

\item Our circles-and-crosses description provides not only a convenient way to visualize parity conservation in the Moore-Read state, but also a way to rigorously construct and study wavefunctions. This is because each root configuration really corresponds to a Jack polynomial, which after appropriate normalization gives a quasihole state. This enables a pathway to probe the microscopic structure of the MR conformal Hilbert space, which will be reported in a future work.
\end{itemize}

\section{Fusion channel selection and braiding statistics from tree diagrams}
\subsection{Fusion Tree Diagram - A very brief introduction}
Here we provide a brief introduction to the fusion tree diagram, which is a convenient tool to discuss the relation between fusion channel selection. Fusion tree diagrams find their application in calculating the explicit first-quantized wavefunctions from conformal blocks. From these wavefunctions, the monodromy of the quasiholes can be read off directly. For the purpose of this paper, it is enough to discuss the fusion tree on their own; for more details on evaluating the fusion matrices and monodromy, see e.g. Ref. \cite{bonderson2008interferometry}. This section does \emph{not} present any new result, but provides an alternate interpretation of the main analysis in the main text that some may find illuminating.

We represent a state with two $\sigma$-anyon with the following diagrams:
\begin{equation}
\label{tree two sigmas}
\centering
\includegraphics[height=36pt]{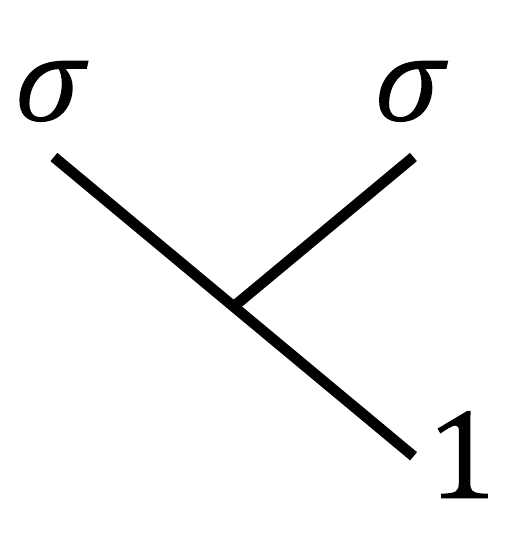}
\text{\hspace{1cm}or\hspace{1cm}}
\includegraphics[height=36pt]{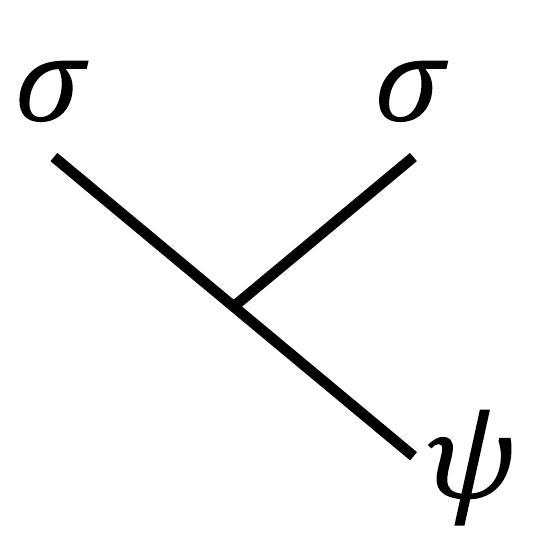}
\end{equation}
The fusion rule $\sigma\times\sigma=1+\psi$ means two diagrams are possible, corresponding to the even sector (left) and odd sector (right) respectively. Each diagram are to be read from bottom to top. Each diagram in Eq.(\ref{tree two sigmas}) represents the process of creating a pair of $\sigma$-anyons from vacuum (``$1$'') or from a single Majorana fermion (``$\psi$''). A four-$\sigma$ state can be similarly represented, respectively in the even and odd sector, by the following diagrams:
\begin{equation}
\label{tree four sigmas}
\centering
\includegraphics[height=64pt]{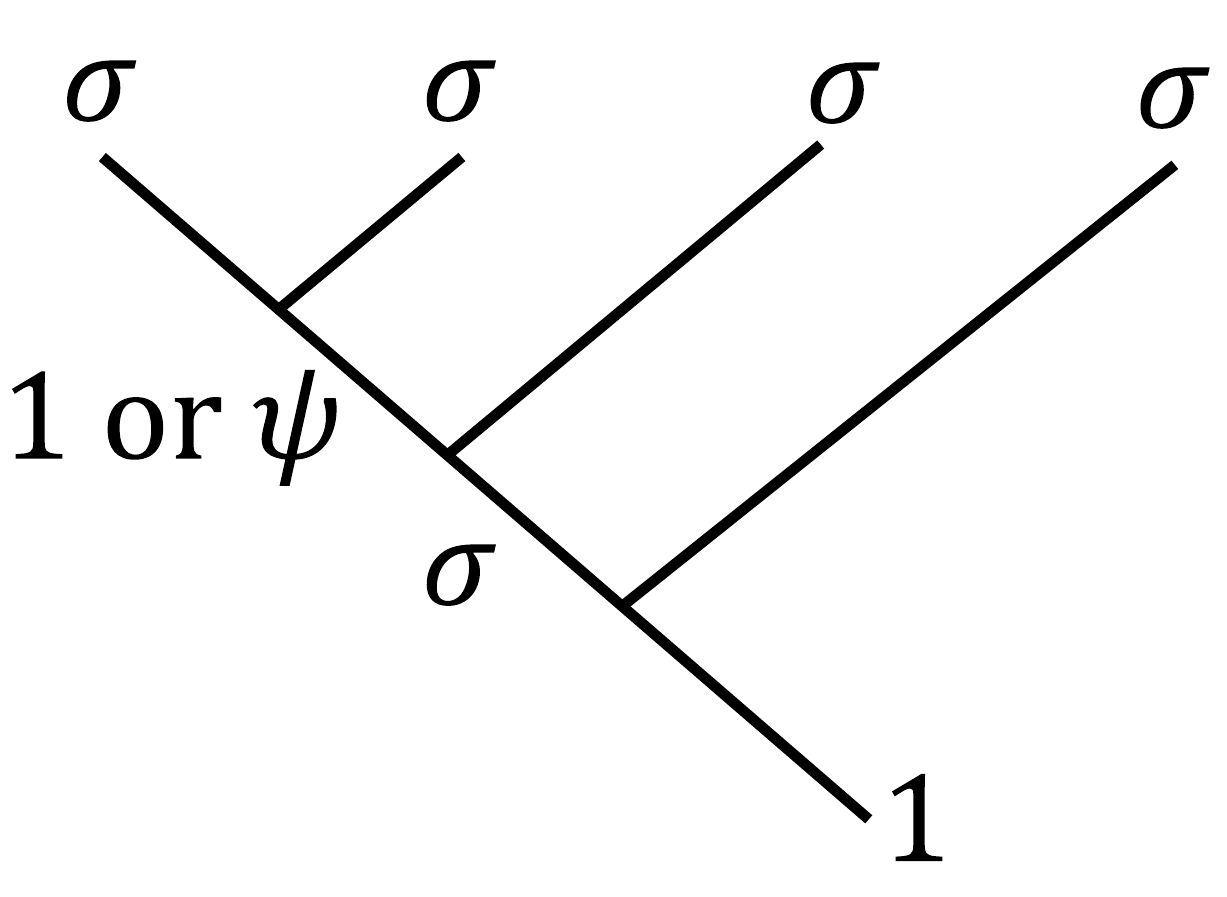}
\text{\hspace{0.5cm}or\hspace{0.5cm}}
\includegraphics[height=64pt]{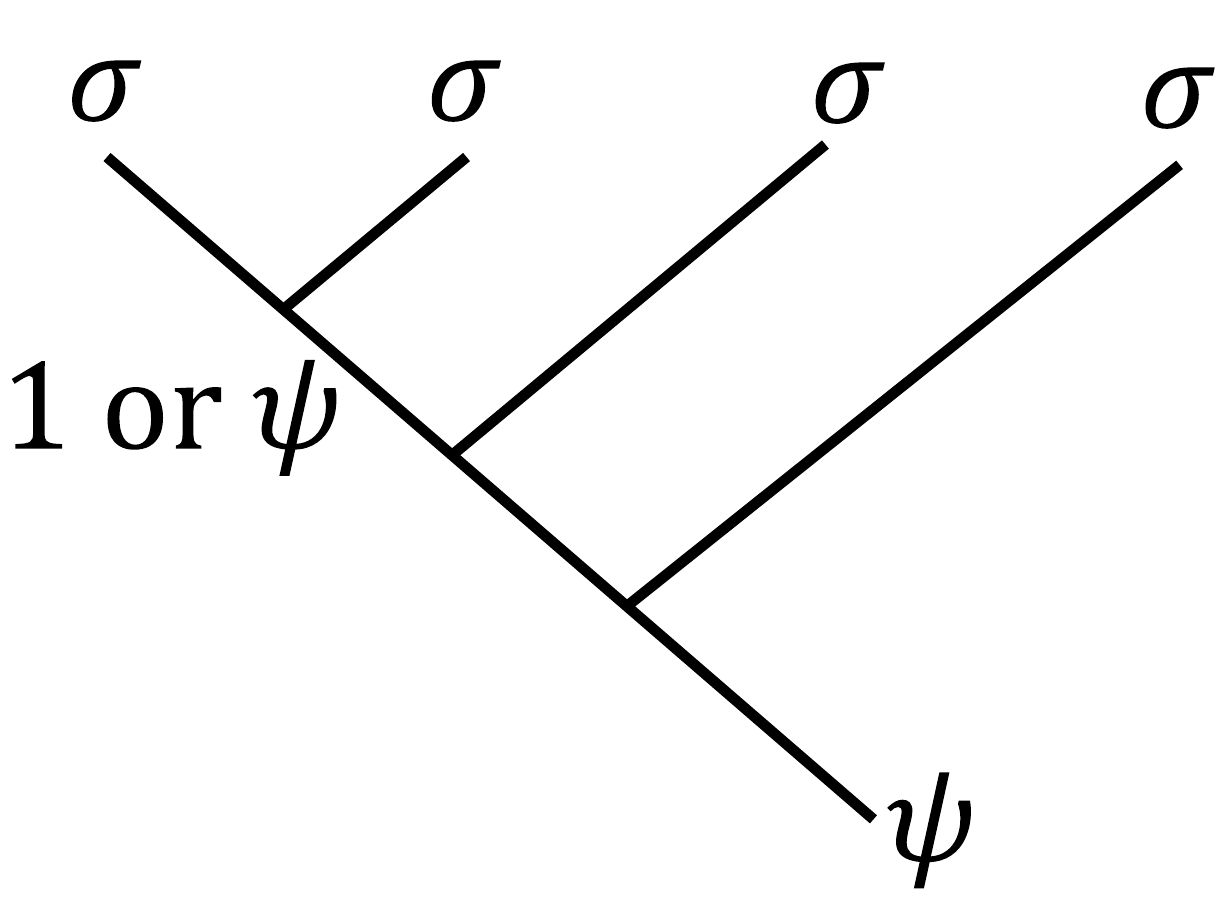}
\end{equation}
Each diagram in Eq.(\ref{tree four sigmas}) represents the creation of four $\sigma$ quasiholes from three pair creation processes (also called ``splitting''). In the second splitting, the fusion rule allows for two possible outcomes, so each diagram actually represents a two-fold degenerate space of four quasiholes. A general diagram representing $2n$ $\sigma$-anyons is as follows 
\begin{equation}
\label{tree 2n sigmas}
\includegraphics[height=128pt]{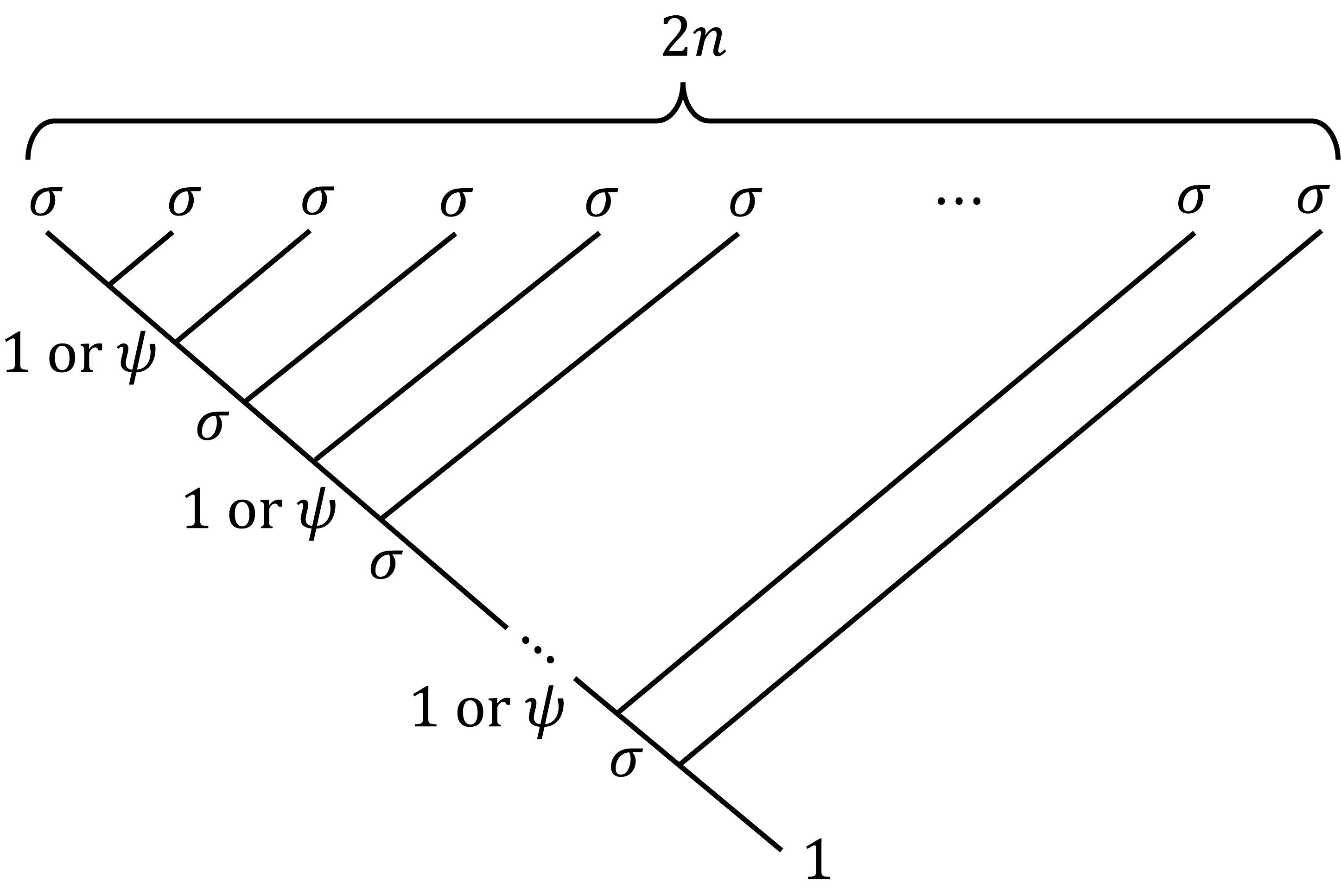}
\end{equation}
for the even sector (for the odd sector, the bottom $1$ is replaced with $\psi$). The degeneracy of this space can also be read off from the diagram: $2n$ quasiholes are created by $2n-1$ processes, $n-1$ of which independently have two possible outcomes, thus the total degeneracy is $2^{n-1}$.
\subsection{Effect of fusion channel selection on anyon statistics}
The effect described above can be seen from the following diagrams:
\begin{align}
\label{tree 4qh fusion}
\includegraphics[height=56pt]{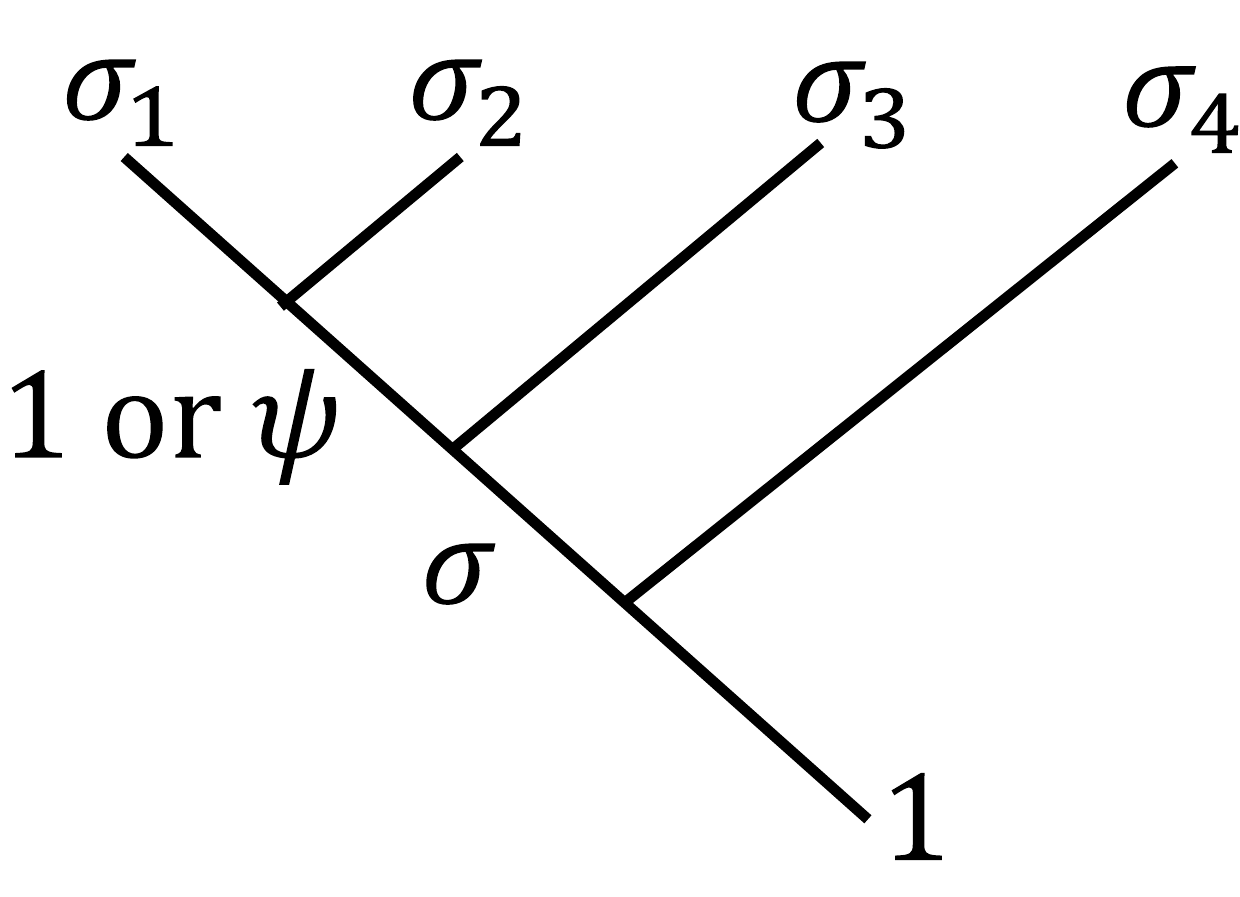}\xrightarrow{\text{fuse $\sigma_3$ and $\sigma_4$}}
\includegraphics[height=56pt]{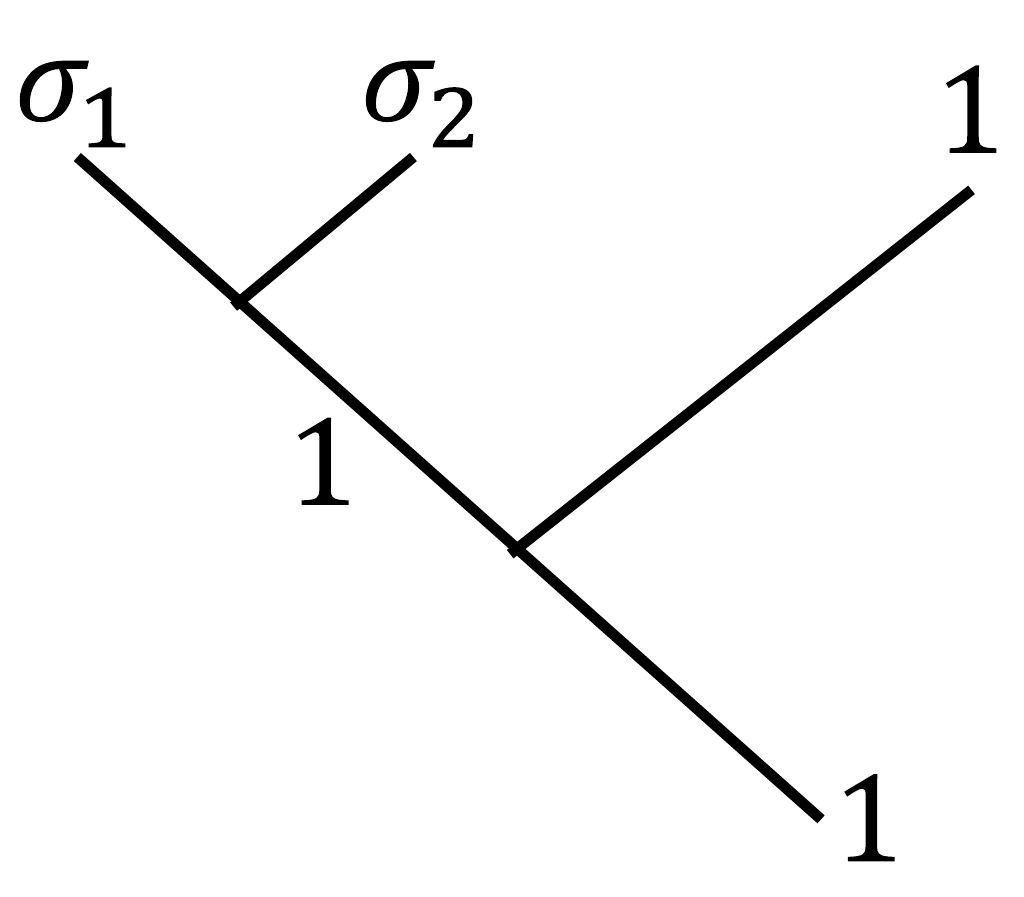}
\text{ or }
\includegraphics[height=56pt]{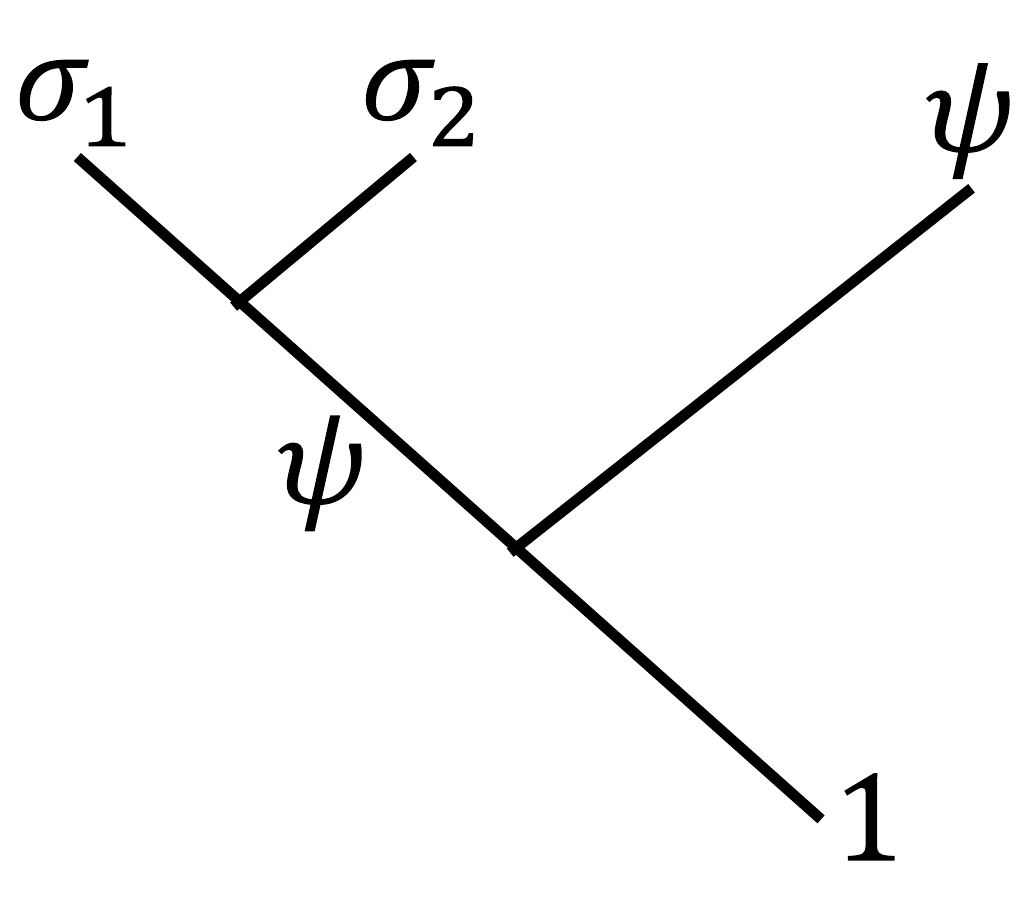}
\end{align}
where after $\sigma_3$ and $\sigma_4$ are fused and their fusion outcome is determined by the dynamics. The remaining $\sigma_1$ and $\sigma_2$ behave like the even sector (if the fusion outcome is $1$) or odd sector (if the fusion outcome is $\psi$). A similar effect can be seen from the diagrams for six quasiholes:
 \begin{align}
\label{tree 4qh fusion}
\includegraphics[height=72pt]{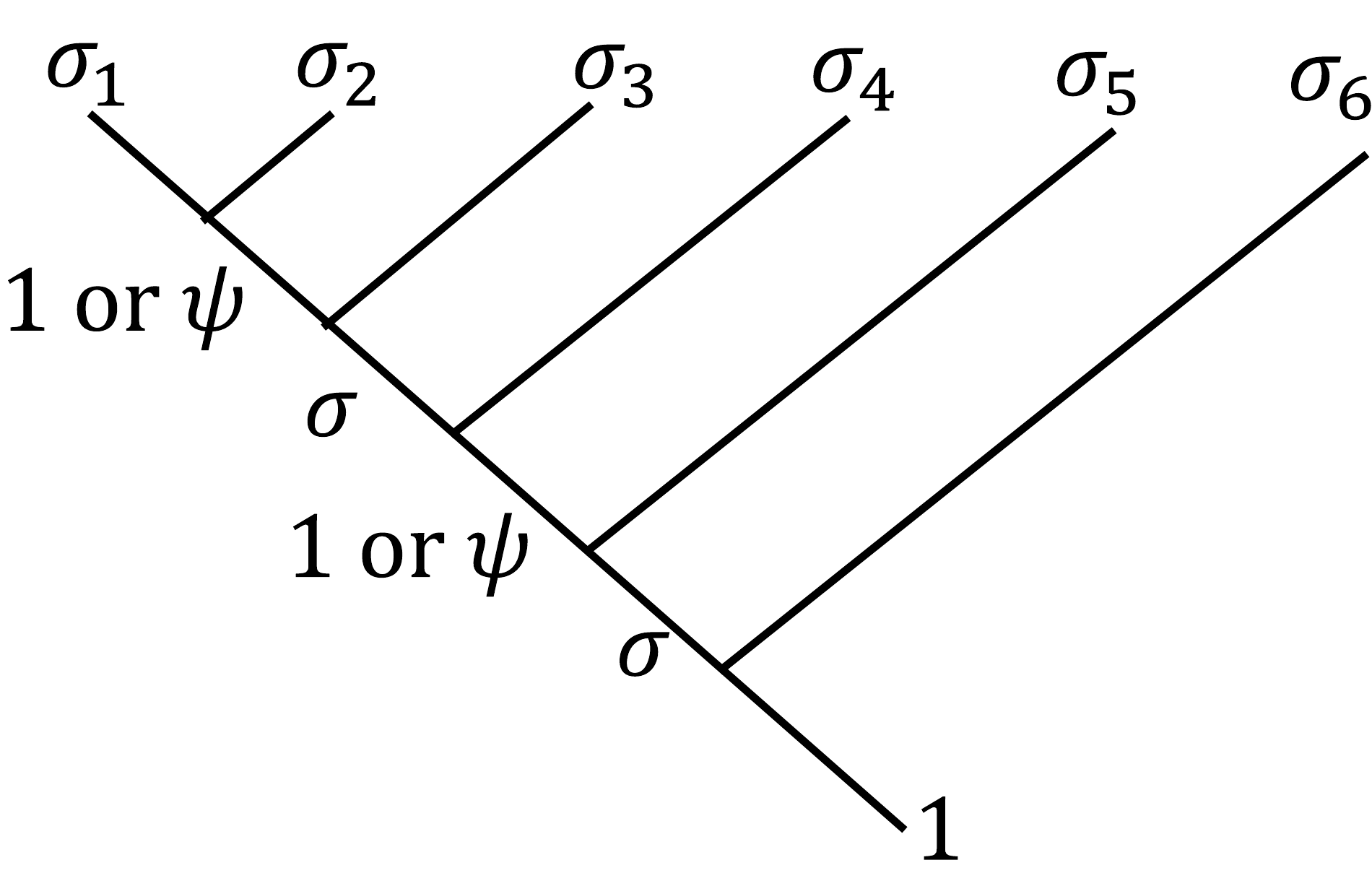}\xrightarrow{\text{fuse $\sigma_5$ and $\sigma_6$}}
\includegraphics[height=72pt]{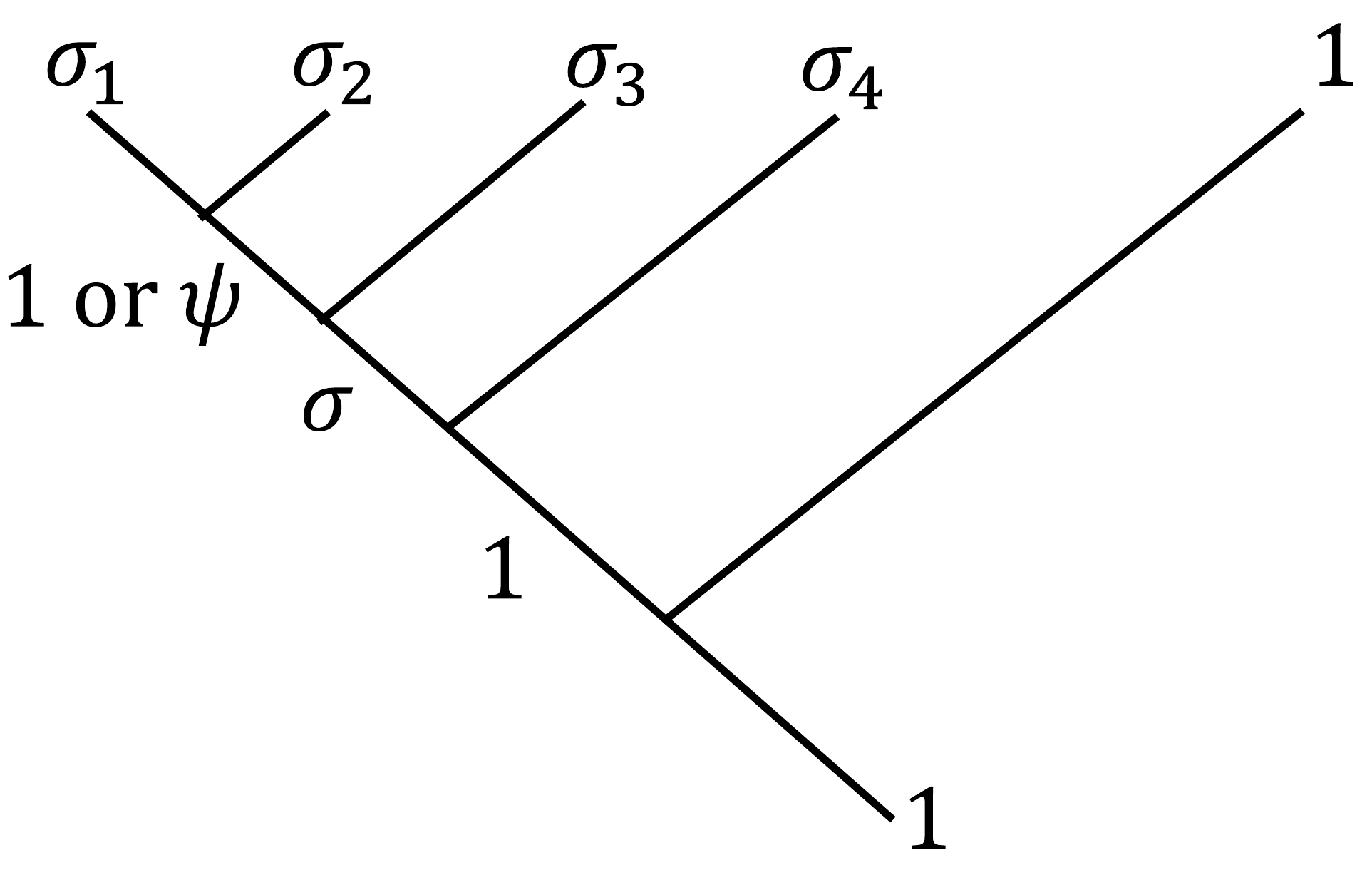}
\text{ or }
\includegraphics[height=72pt]{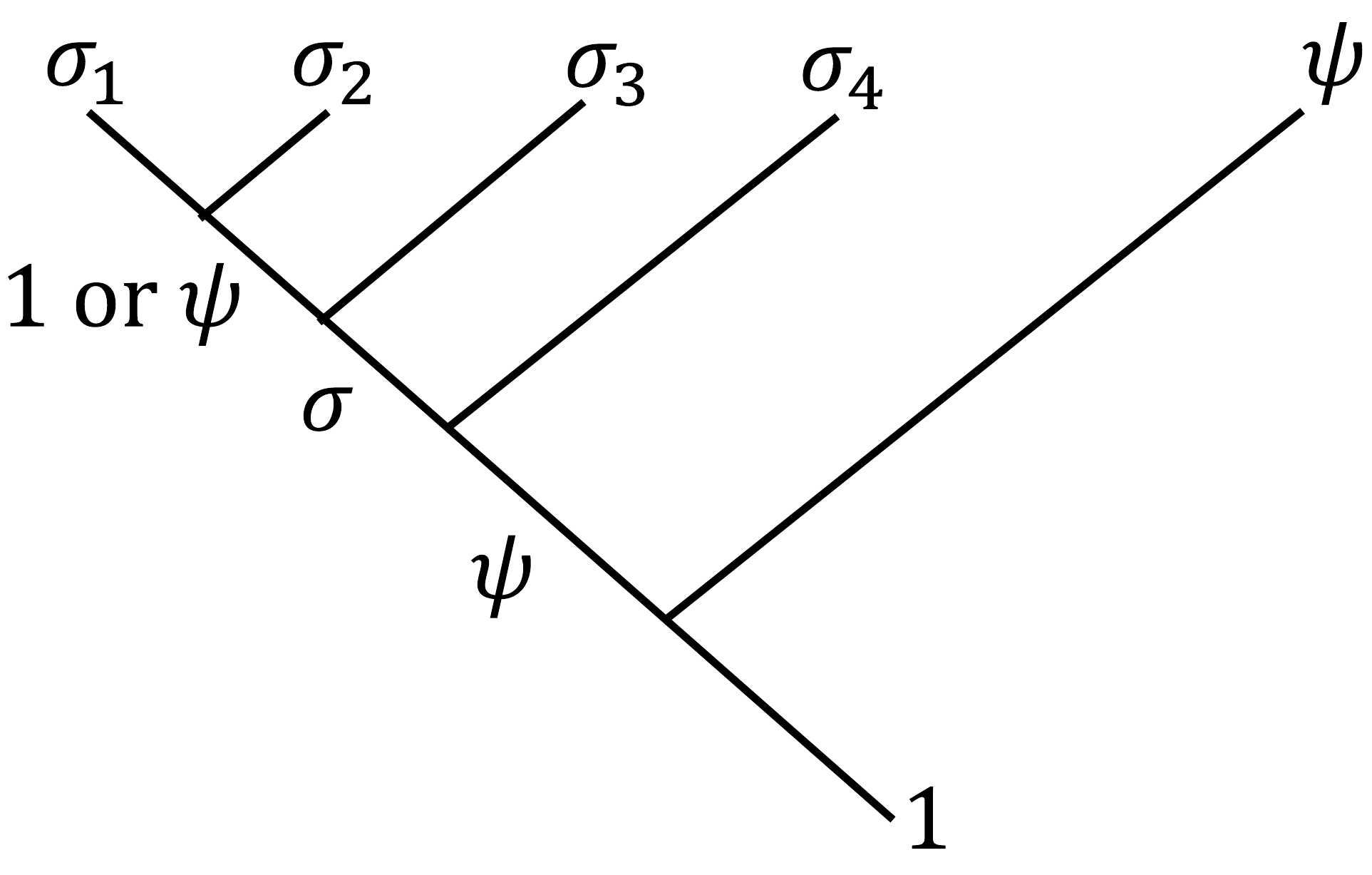}
\end{align}
Here, after the fusion of $\sigma_5$ and $\sigma_6$, the ground state manifold consisting of $\sigma_1$, $\sigma_2$, $\sigma_3$, $\sigma_4$, and the fused quasihole is still two-fold degenerate. Thus, the four $\sigma$-anyons still exhibit non-Abelian statistics regardless of the fusion outcome preferred by the dynamics. However, due to the different ``sub-diagrams'', the two fusion outcomes will results in different Berry matrices\cite{Bonderson2011}.
\section{Creation energy of MR quasiholes}
%\textcolor{blue}{I will use all the exact figure and equation numbers in your main content if they are referred to.}

\begin{figure*}
    \centering
    \includegraphics[width=1\textwidth]{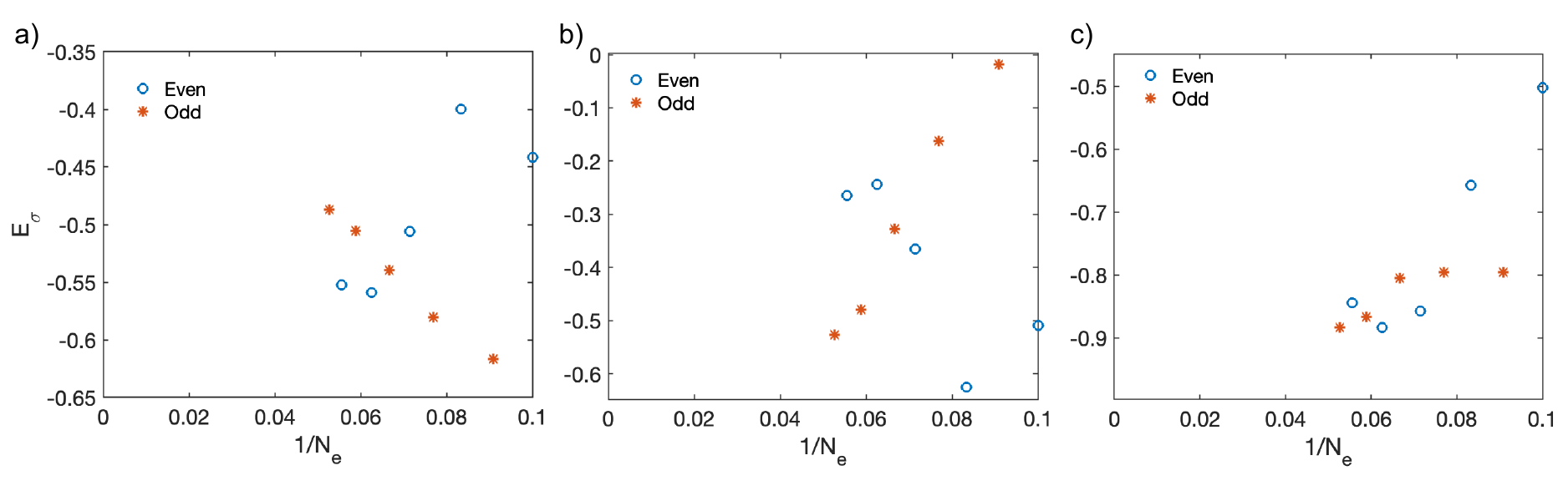}
    \caption{Energy difference between the NN state and the NS state with different pseudopotentials: (a) with $\hat{V}_1$, (b) with $\hat{V}_3$, (c) with $\hat{V}_5$. The systems' sizes are from 10 electrons to 18 electrons, increased every time by two for even and from 11 electrons to 19 electrons increased every time by two for odd.}
    \label{fig_mr_pseudo}
\end{figure*}

The self-energy of $\sigma$-anyons can never be determined directly because of the simultaneous creation of two $\sigma$-anyons from one flux insertion in MR non-Abelian phases. But it can be asymptotically computed by the method we shall explain here. The expectation value of any electron-electron interaction regarding the state with two $\sigma$-anyons has three components: the ground-state energy, two quasihole creation energy (effective self-energy), and the interaction between the two quasiholes. Considering the state represented by Eq. (8) where two $\sigma$-anyons are positioned at the North and South pole, the separation between them ($2R$ as $R$ being the radius of the sphere) is enlarged when we increase the system's size. When $R$ (namely the separation between two anyons) is large enough, there should be no interaction between them and only self-energies remain. The energies of the state in Eq. (8) with the model Hamiltonian $\hat{V}_1$, $\hat{V}_3$, and $\hat{V}_5$ subtracted by the ground-state energies are shown in Figure \ref{fig_mr_pseudo} for both odd and even numbers of electrons. Though the anyon types are both $\sigma$, their effective self-energies in the two different sectors are different at least up to the finite system sizes we could obtain ($18$ electrons for $1$-sector and $19$ electrons for $\psi$-sector).

\section{Layer Thickness and Screening as Tuning Parameters}
In the main text, we discussed the method of preserving the non-Abelian degeneracy by tuning the relative strengths of the pseudopotentials $\hat V_3^{2bdy}$ and $\hat V_1^{2bdy}$. Roughly speaking, $\hat V_1^{2bdy}$ is a very short-ranged interaction while $\hat V_3^{2bdy}$ is slightly longer-ranged. Thus, we translate the idea proposed with the toy model in the main text into real experimental setups by proposing that the range of the effective electron-electron interaction can be used as a tuning parameter to balance the two fusion channels. There are two possible ways of tuning the interaction range: by screening (which effectively decreases the range) and by increasing the sample thickness (which effectively increases the range). Here we provide a brief discussion of these interactions.

\subsection{Finite thickness effect}
To model finite layer thickness we use the Zhang-Das Sarma (ZDS) interaction\cite{zhang1986excitation}:
\begin{equation}
V_{\text{ZDS}}(\mathbf{r}) = \frac{1}{\sqrt{|\mathbf{r}|^2 + d^2}}, \ \  V_{\text{ZDS}}(|\mathbf{q}|) = \frac{e^{-d |\mathbf{q}|/2}}{|\mathbf{q}|}
\end{equation}
Besides ZDS interaction, there are two more realistic interactions considering finite width effect: 
Fang-Howard interaction $V_{\text{FH}}$ used in heterostructures and infinite square-well potential $V_{\text{SQ}}$ suitable for 2D GaAs which are 

\begin{equation}
V_{\text{FH}}(|\mathbf{q}|) = \frac{9}{8|\mathbf{q}|}\frac{24+9d|\mathbf{q}|+(d|\mathbf{q}|)^2}{(3+d|\mathbf{q}|)^3},
\label{FH_potential}
\end{equation}
\begin{equation}
V_{\text{SQ}}(|\mathbf{q}|) = \frac{1}{|\mathbf{q}|} \frac{3d|\mathbf{q}|+\frac{8\pi^2}{d|\mathbf{q}|} - \frac{32\pi^4(1-e^{-d|\mathbf{q}|})}{ (d|\mathbf{q}|)^2 [(d|\mathbf{q}|)^2 + 4\pi^2] } }{ (d|\mathbf{q}|)^2 + 4\pi^2 }.
\label{SQ_potential}
\end{equation}
The larger $d$ is, the coefficients in the pseudopotential decompositions of these three interactions become closer. When $d=0$, these interactions degenerate to Coulomb interaction again. However, based on our studies, the results of the ZDS interaction can largely imitate the other two more realistic interactions, and there is no qualitative difference.

The effect of layer thickness on the creation energy of the two types of charge-$e/2$ quasiholes are shown in Fig. \ref{fig_variational-energy}a. In general, the difference in their energies changes non-monotonously with increasing layer thickness. This is understandable, as increasing the layer thickness increases the range of the effective interaction, which increases not only the amount of $\hat V_3^{2bdy}$ (relative to $\hat V_1^{2bdy}$) but also the amount of $\hat V_5^{2bdy}$, $\hat V_7^{2bdy}$, and so on. Out of these pseudopotentials, only $\hat V_3^{2bdy}$ enegetically favors the $1$-type quasihole. If the thickness is too small, $\hat V_3^{2bdy}$ is not comparable to $\hat V_1^{2bdy}$, but if the thickness is too large, $\hat V_3^{2bdy}$ is also suppressed by other terms $\hat V_m^{2bdy}$ for $m\geq 5$. Thus, in both limits we may expect the interaction to strongly favor the $\psi$-type quasihole. However, in the intermediate regime we find a range of energy difference which suggests that layer thickness can still be a useful parameter (especially when the full Hamiltonian contains other terms, such as one-body potentials, that favors the $1$-type quasiholes). 

\subsection{Screening effect}
The screened Coulomb interaction is modelled by the Yukawa potential
\begin{equation}
\label{Yukawa}
V_{\text{Yukawa}}(\mathbf r_i) = \sum_{i<j}\frac{e^{-\lambda|\mathbf r_i-\mathbf r_j|}}{|\mathbf r_i-\mathbf r_j|} 
\end{equation}
where $\lambda$ is the screening strength (inverse of screening length). 

\begin{figure}
\centering
\includegraphics[width=0.5\linewidth]{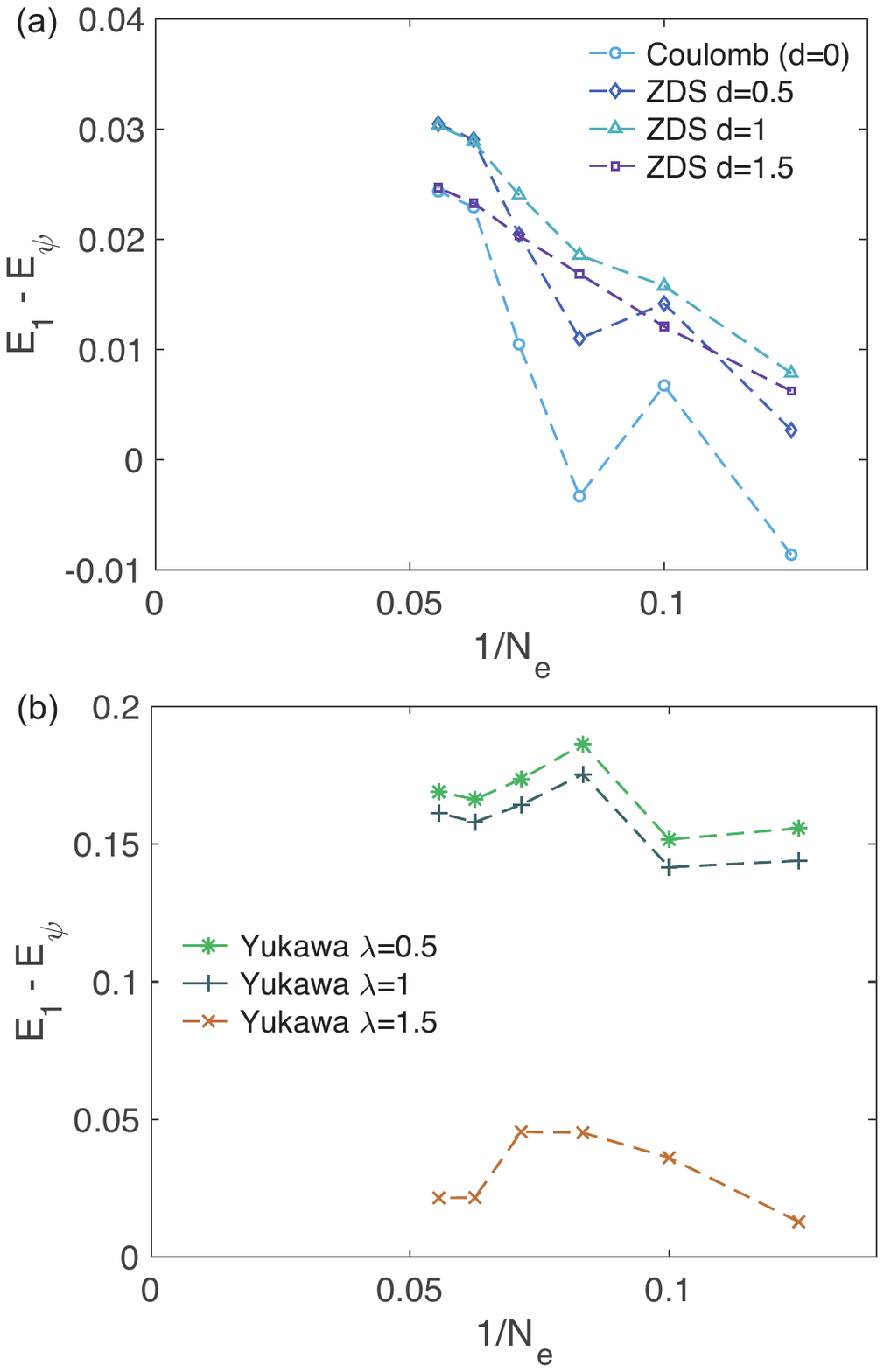}
\caption{(a) Finite-size scaling of the difference in self-energy of the two types of charge-$e/2$ anyons in systems with different thickness $d$ modelled by the ZDS interaction. (b) Finite-size scaling of the difference in self-energy of the two types of charge-$e/2$ anyons under screened Coulomb interaction with different screening strengths, modelled by the Yukawa interaction. Similar to Fig. \ref{fig_self-energy} in the main text, a positive energy difference means the $\psi$-type quasihole is energetically favorable.}
\label{fig_variational-energy}
\end{figure}

The effect of screening on quasihole creation energy is shown in Fig. \ref{fig_variational-energy}b for a limited range of parameter. Here we find a stark contrast between smaller screening ($\lambda\leq 1$) and strong screening ($\lambda=1.5$). While in both cases, $\psi$-type quasiholes are energetically favorable, the energy difference is decreased by one order of magnitude in the latter case compared to the former. This is because stronger screening suppresses the effect of higher pseudopotential terms ($\hat V_m^{2bdy}$ with $m\geq 5$). As a result, the resulting effective interaction only contains $\hat V_1^{2bdy}$ and $\hat V_3^{2bdy}$. The fact that the energy difference (brown line in Fig. \ref{fig_variational-energy}b) is very small implies that potentially this choice of screening strength is close to the ideal balance between these two pseudopotential where non-Abelian degeneracy is preserved (discussed in the main text). Once again, the exact energetics also depend on other terms in the full Hamiltonian such as the one-body potentials, but the drastic change seen in Fig. \ref{fig_variational-energy}b means screening strength can be an effective tuning parameter for favoring the $\psi$-type quasihole, countering the effect of one-body potential, which favors the $1$-type.

\section{Details on the numerical methods}
Here we discuss the details of the braiding phase numerics presented in the main text. We start by describing the toy Hamiltonian used and discuss its spectrum. Afterward we recall the method for numerically calculating the Berry phase and extracting the braiding statistics, first reported in Ref. \cite{trung2023spin}.

\subsection{General Hamiltonian and numerical constrains}
The general Hamiltonian we use to numerically simulate the braiding process is
\begin{align}
\hat H(t) &= \lambda_0 \hat V_3^{3bdy}+\hat H_1(t)\label{H}\\
\hat H_1(t) &= \lambda_1 \hat V_1^{2bdy}+ \hat v_0V_{\text{pins}}(t)\label{H_1}
\end{align}
where $\hat V_m^{nbdy}$ denotes the $m$-th $n$-body pseudopotential, $\lambda_0$ and $\lambda_1$ are some scalars, and $\hat V_{\text{pins}}$ is a one-body pinning potential that depends on some fictituous ``time'' parameter $t$. The exact form of $\hat V_{\text{pins}}$ will be described later.

We take the limit $\lambda_0\to\infty$, ensuring that all low-lying (finite energy) states reside within the Moore-Read conformal Hilbert space (MR CHS). A basis for this Hilbert space can be obtained by taking all the Jack polynomials and orthonormalize them (in general, Jack polynomials form a non-orthogonal but complete basis for the MR CHS).

The pinning potential $\hat V_{\text{pins}}$ is based off the following one-body potential:
\begin{equation}
\label{one-body}
\hat V_0^{1bdy}(\theta_0,\phi_0) = |\theta_0,\phi_0\rangle\langle\theta_0,\phi_0|
\end{equation}
where $|\theta_0,\phi_0\rangle$ is the coherent state centered at $(\theta_0,\phi_0)$ on the sphere (here $\theta$ denotes the azimuthal angle and $\phi$ denotes the polar angle). Its first quantized wavefunction can be expressed in terms of the (unnormalized) lowest Landau level eigenstate $\varphi_m(u,v)= u^{S+m}v^{S-m}$ as
\begin{equation}
\label{coherent state}
\psi_{\theta_0,\phi_0}(u,v)\propto\sum_{m}u_0^{S-m}v_0^{m}\varphi_m(u,v)
\end{equation}
where 
\begin{align}
u&=\cos\left(\theta/2\right)e^{i\phi/2}\\
v&=\sin\left(\theta/2\right)e^{-i\phi/2}
\end{align}

In the numeric presented in the main text, a three-pin configuration was used:
\begin{equation}
\label{pin configuration}
\hat V_{\text{pins}}(t) = \hat V_0^{1bdy}(\theta_0,\phi_0(t))+\hat V_0^{1bdy}(\theta_0,\phi_0(t)+\pi)+2\hat V_0^{1bdy}(\pi,0)
\end{equation}
The system consists of two potential pins positioned opposite of each other on a small circle parallel to the equator (first two terms) and a third pin at the south pole (third term) (see inset of Fig. 7a in the main text). The pin at the south pole is twice the strength of each of the other two pins, so that when $\theta_0=0$ it is equivalent to two identical pin at the north and south poles. The polar angle $\phi_0(t)$ is varied between $[0,\pi]$ from $t=0$ to $t=T$ so that the two pins are exchanged at the end of the process.

In finite-size numerics, the choice of $\lambda_1$ parameter in Eq.\ref{H_1} is also important. Ideally, we would like to compare the difference between two case, $\lambda_1=0$, where $1$-anyon fusion channel is energetically favored by $\hat V_{\text{pins}}$, and large $\lambda_1$, where $\psi$-anyon fusion channel is preferred. However, at finite quasihole separation, $\hat V_1^{2bdy}$ affects not only the creation energy of the three types of anyons, but also the interaction energy between them. In general, the interaction energy varies non-monotonously as a function of anyon separation\cite{xu2024}. Thus, the presence of $\hat V_1$ in a finite system greatly complicates the dynamics of the system under study.

In our numerics, we empirically choose $\lambda_1=5.0$, which is large so that the spectrum is qualitatively different (see Fig. 6 in the main text) but not so large that the quasihole interaction dominates. As pointed out in the main text, there exist more than two states with low-lying energy. The two lowest energy states are the states of interest---consisting of two charge-$e/4$ anyons and one charge-$e/2$ anyons of either $1$-type or $\psi$-type; the remaining are various other anyon configurations. Especially at finite temperature, all of these states may be important for understanding experimental outcomes. However, their nature remains to be understood.

\subsection{Berry phase and braiding phase}
The Berry phase gained by a state $|\psi(t)\rangle$ during a process from $t=0$ to $t=T$ is calculated numerically by computing the following quantity
\begin{equation}
\label{Berry}
\mathcal D e^{i\gamma}=\langle\psi(t_0)|\psi(t_1)\rangle\langle\psi(t_1)|\psi(t_2)\rangle\langle\psi(t_2)|...|\psi(t_{N-1})\rangle\langle\psi(t_{N-1})|\psi(t_N)\rangle\langle\psi(t_N)|\psi(t_0)\rangle
\end{equation}
where the $t_i$'s are discretized time: $0=t_0<t_1<t_2<...<t_N=T$. In the limit $N\to\infty$, $\mathcal D\to 1$ and $\gamma\to\gamma_{\text{Berry}}$ is the desired Berry phase. In practice, $N$ is chosen large enough such that $\mathcal D$ is close to unity, and the closeness of $\mathcal D$ to unity indicates how accurate the determined value of $\gamma$ is to the real value $\gamma_{\text{Berry}}$. Eq. (\ref{Berry}) also has an advantage of being gauge-independent. This is specially useful for numerics where $|\psi(t_i)\rangle$ is obtained by exact diagonalization (ED), since any ED algorithm can only determine the eigenvector up to a random phase factor. However, we can see in Eq. (\ref{Berry}) each ket has a corresponding bra, thus cancelling out the effect of such random phases.

The braiding phase can be extracted from the Berry phase data. In general, when a quasihole moves in a closed circle on the sphere, there exists three contributions to the Berry phase: the Aharonov-Bohm phase, the parallel transport phase (coming from the curvature of the sphere), and the braiding phase (coming from any other quasihole presence within the closed loop). The first two terms are both proportional to the solid angle $\Omega$ enclosed by the loop and only the braiding phase is $\Omega$-independent. Thus, we can extract the braiding phase by determining the $\Omega$-independent part of the phase by linear fitting. There is however a caveat: since quasiholes are objects with a finite size, this is only true if they are sufficiently far apart from each other. Thus, in practice, in a plot of the Berry phase against $\Omega$, we find that only the portion of the graph at larger $\Omega$ can be fitted with a line (see Fig. 7 in the main text). However, in most cases with larger systems, the braiding phase can be reliably extracted from the $y$-intercept of the linear fit of the right portion of the graph.

\section{Tuning the non-Abelian gap in realistic systems}
\subsection{Balancing $\hat V_1^{2bdy}$ and $\hat V_3^{2bdy}$}
\begin{figure}
\centering
\includegraphics[width=0.7\linewidth]{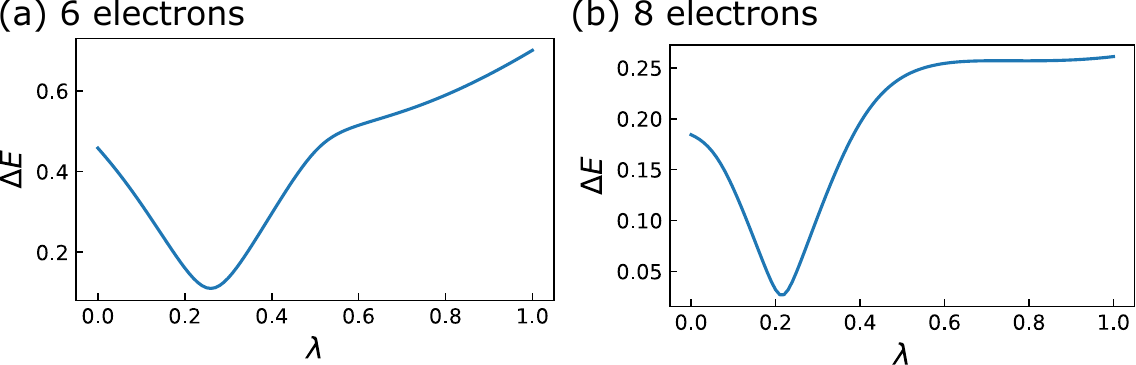}
\caption{The non-Abelian gap as a function of interpolation between $\hat V_1^{2bdy}$ ($\lambda=0$) to $\hat V_3^{2bdy}$ ($\lambda=1$) (see Eqs. (\ref{general}) and (\ref{v1v3 interpolate})) for (a) $N_e=6$ electrons and (b) $N_e=8$ electrons.}
\label{fig_nonAbelian_gap_sizes}
\end{figure}
As discussed in the main text, the non-Abelian gap refers to the amount of energy required for a pair of $\psi$-anyons to spontaneously transmute to a pair of $1$-anyons. Ensuring that this energy is zero (or as low as possible) is crucial in observing non-Abelian statistics. In numerics, we study this gap by working on a MR system with four quasiholes in the even sector and apply a two-body interaction as well as two pinning potentials:
\begin{equation}
\label{general}
\mathcal H = \lambda_0 \hat V^{3bdy}+\hat V^{2bdy} + v_0\hat V_{\text{pins}}
\end{equation}
Setting $\lambda_0\to\infty$ ensures the system stays in the MR CHS, while two potential pins specified by $\hat V_{\text{pins}}$ ensures only two quasiholes (each with charge $e/2$) are present in the system. If the two pins are far enough from each other (one at each pole on the sphere), we expect the ground state to be either two $1$-type quasiholes or two $\psi$-type quasiholes, depending on the exact form of the two-body interaction $\hat V^{2bdy}$. Since $\hat V_1^{2bdy}$ and $\hat V_3^{2bdy}$ each prefers a different type of quasiholes (namely, the $\psi$-type and $1$-type, respectively), it is enough to consider a toy model that interpolates between the two:
\begin{equation}
\label{v1v3 interpolate}
\hat V^{2bdy}=(1-\lambda)\hat V_1^{2bdy} + \lambda \hat V_3^{2bdy}
\end{equation}
In general, one expect the gap between the two lowest energy to close as $\lambda$ is tuned from 0 to 1, as the ground state transmutes from two $\psi$-anyons into two $1$-anyons. This is what we see in Fig 8 of the main text for $N_e=10$ electrons. Here we present more numerical results to show that this feature is robust across many system sizes (see Fig. \ref{fig_nonAbelian_gap_sizes}). While the exact value of $\lambda$ at which the gap attains a minimum value changes, it is true across all system sizes that there always exists an interaction interpolating between $\hat V_1^{2bdy}$ and $\hat V_3^{2bdy}$ where the non-Abelian gap is minimized.

\subsection{In presence of impurities}
\begin{figure}
\begin{center}
\includegraphics[width=\linewidth]{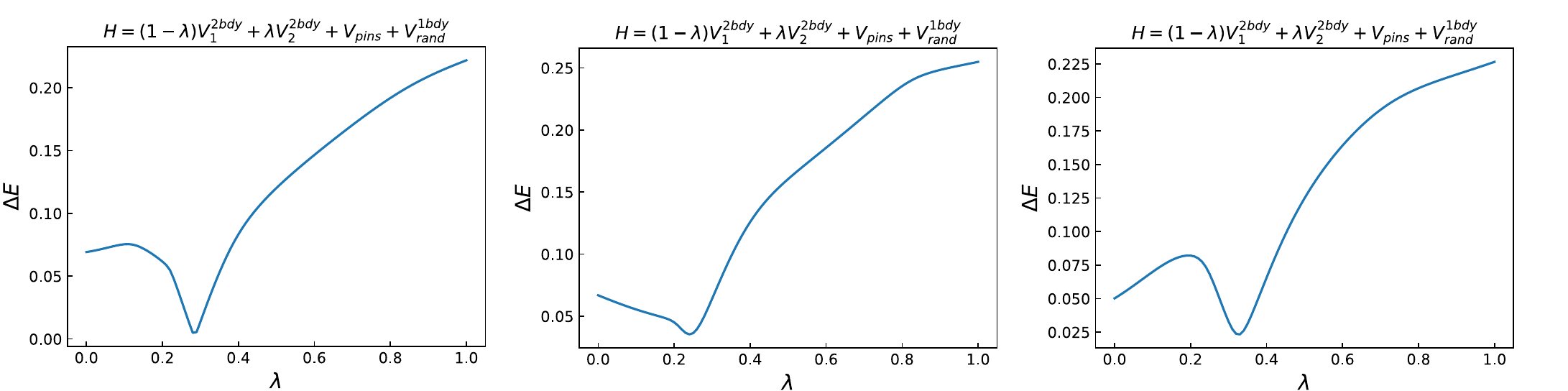}
\caption{The energy difference between the lowest two eigenvalues of Eq. (\ref{Hamiltonian with impurities}) for three different random impurity potential profiles. Calculated for a system with 8 electrons and 16 orbitals.}
\label{fig:impurity}
\end{center}
\end{figure}

In this section we briefly discuss a more realistic scenario where some impurities are introduced to the system. We model impurities using a random one-body interaction $\hat V_{\text{rand}}^{1bdy}$:
\begin{equation}
\label{random one-body}
\hat V_{\text{rand}}^{1bdy}=\sum_{m_1,m_2}\eta_{m_1m_2}\hat c^\dagger_{m_1}\hat c_{m_2}
\end{equation}
where $\hat c^\dagger_m$ and $\hat c_m$ respectively creates and annihilates an electron at orbital $m$, and $\eta_{m_1m_2}$ are complex numbers whose real and imaginary parts are each a random number between $-1$ and $1$, with a constraint that $\eta_{m_2m_1}=\eta_{m_1m_2}^*$. The overall strength of this random potential is modulated by a real scalar factor $v_i$ in Eq. (\ref{Hamiltonian with impurities}) below. In general, this overall factor cannot be too large, otherwise the random potential will dominate over any interaction in the system. The total toy-model Hamiltonian used in this section is
\begin{equation}
\label{Hamiltonian with impurities}
\mathcal H = \lambda_0 \hat V^{3bdy}+(1-\lambda) \hat V_1^{2bdy} + \lambda\hat V_3^{2bdy} + v_0\hat V_{\text{pins}} + v_i\hat V_{\text{rand}}^{1bdy}
\end{equation}

We see in Fig. \ref{fig:impurity} that no matter the profile of the impurities (which is generated at random), by tuning $\lambda$ there is always a range within which the gap reaches a minimum value which is very small. In general, this result is not very trivial, as the effect of a random potential is not easily summarized in a manner similar to Table 1 in the main text. In general, it depends on the positions and sizes of the local maxima and minima in the protential profile and also on the two-body interaction of the system. However, from our numerical simulation we see empirically that as long as the overall strength of the impurity is not too big, tuning the two-body interaction remains an effective method of tuning the non-Abelian gap.

A random potential on the sphere completely breaks rotational symmetry, and thus our numerics must be done on all $L_z$ sectors. This results in a much larger Hilbert space compared to the rotationally symmetric case with only the $L_z=0$ sector involved. As a result, the largest system size accessible is decreased. However, we expect this result to be consistent accross all system sizes.

\subsection{Sample thickness and screening as tuning parameters}
Using the same principle, we can also investigate the effect of sample thickness and screening in tuning the non-Abelian gap. As suggested earlier, increasing the thickness increases the effective interaction range, thereby effectively increasing the $\hat V_3^{2bdy}$-to-$\hat V_1^{2bdy}$ ratio. On the other hand, increasing the screening strength reduces the effective interaction range, thereby effectively reducing this ratio. However, any realistic interaction contains not only $\hat V_1^{2bdy}$ and $\hat V_3^{2bdy}$, but also higher terms $\hat V_5^{2bdy}$, $\hat V_7 ^{2bdy}$, and so on. These terms may have unpredictable effect on the anyons dynamics.

Here we study the effect of these parameter in tuning the non-Abelian gap by the following toy Hamiltonian
\begin{equation}
\label{Hamiltonian with real itr}
\mathcal H = \lambda_0 \hat V^{3bdy}+ \hat V^{2bdy} + v_0\hat V_{\text{pins}} + v_i\hat V_{\text{rand}}^{1bdy}
\end{equation}
where $\hat V^{2bdy}$ is either the ZDS interaction $\hat V_{\text{ZDS}}^{\text{(n)}}$ or the Yukawa potential $\hat V_{\text{Yukawa}}^{\text{(n)}}$. Here, the superscript (n) denotes that these interactions are \emph{normalized}, i.e. they are scaled by a factor such that all pseudopotential coefficients sum up to 1. This is to ensure that any decrease in the gap is not due to the decrease in overall strength of the interaction. 

\begin{figure}
\centering
\includegraphics[width=0.7\linewidth]{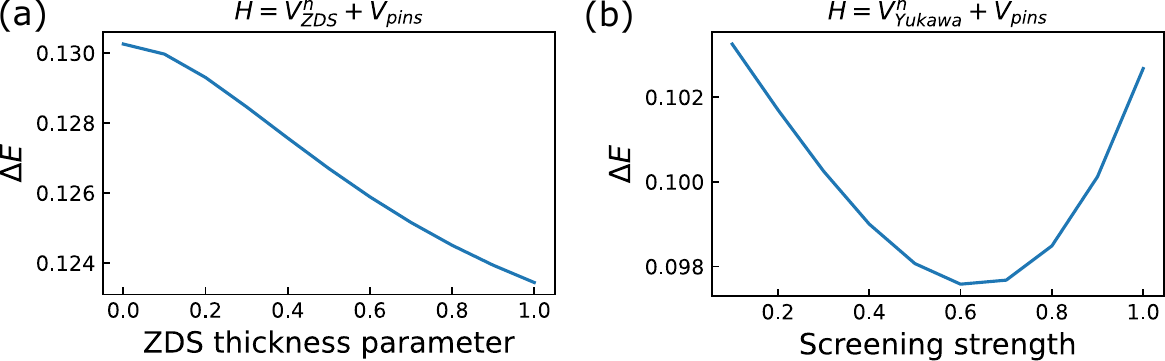}
\caption{(a) The non-Abelian gap as a function of layer thickness with two-body interaction modelled by the ZDS interaction (b) The non-Abelian gap as a function of screen strength with the screened Coulomb interaction modelled by the Yukawa potential. Both are calculated for systems with $N_e=6$ electrons.}
\label{fig_nonAbelian_gap}
\end{figure}

The results for preliminary calculations with $N_e=6$ electrons are shown in Fig. \ref{fig_nonAbelian_gap} which shows the non-Abelian gap decreasing with layer thickness, as expected from increasing the interaction range. On the other hand, the non-Abelian gap does not vary monotonously with screening strength. It does not close exactly, but there is a screening strength at which it achieves a minimum. While these results show a proof of concept that it is possible to tune the non-Abelian gap with experimental parameter, to realize this idea requires further extensive numerics with experimental input. An important thing to note that in order to observe non-Abelian statistics, any interaction that minimizes non-Abelian gap must not also destroy the FQH plateau (this is not a concern in our numerics because we have set $\lambda_0\to\infty$ in Eq. (\ref{Hamiltonian with real itr})).

\bibliography{reb}

\end{document}